\documentclass[aip,superscriptaddress,amsmath,amssymb,pof]{revtex4-1}
\usepackage{natbib}
\usepackage{amsmath}
\usepackage{dcolumn}
\usepackage{bm}
\usepackage{hyperref}
\usepackage{float}
\usepackage{tabularx}
\usepackage{color}
\usepackage{tikz}
\usetikzlibrary{arrows, petri, topaths}
\usepackage{hyperref}
\usepackage{colortbl}
\hypersetup{
colorlinks,
citecolor=blue,
filecolor=blue,
linkcolor=blue,
urlcolor=blue}

\definecolor{Gray}{gray}{0.90}
\newcolumntype{a}{>{\columncolor{Gray}}c}
\definecolor{LightCyan}{rgb}{0.88,1,1}

\begin{document}
\title{Flow reversals in  turbulent convection with free-slip walls}
\author{Mahendra K. Verma}
\email{mkv@iitk.ac.in}
\author{Siddhesh C. Ambhire}
\author{Ambrish Pandey}
\affiliation{Department of Physics, Indian Institute of Technology, Kanpur-208016, India}
\date{\today}

\begin{abstract}
We perform numerical simulations of turbulent convection for infinite Prandtl number with free-slip walls, and study the dynamics of flow reversals. We show  interesting correlations between the flow reversals and the nonlinear interactions among the large-scale flow structures represented by the modes $(1,1), (2,1), (3,1)$ and some others. After a flow reversal, the odd modes, e.g. $(1,1), (3,1)$, switch sign, but the even modes, e.g. $(2,2)$, retain their sign. The mixed modes $(1,2)$ and $(2,1)$ fluctuate around zero.  Using the properties of the modes and their interactions, we show that they form a Klein four-group $Z_2 \times Z_2$.   We also show that for the free-slip boundary condition, the corner rolls and vortex reconnection are absent during a flow reversal, in contrast to active role played by them in flow reversals for  the no-slip boundary condition. We argue that the flow reversals with the no-slip and free-slip boundary conditions are different because they are induced  by nonlinearities $ ({\bf u} \cdot \nabla) {\bf u}$ and $({\bf  u} \cdot \nabla) \theta$ respectively.
\end{abstract}

\maketitle

\section{Introduction}
The velocity field in turbulent convection reverses in random intervals.~\cite{Cioni:JFM1997, Niemela:Nature2000, Brown:PRL2005, Brown:JFM2006, Mishra:JFM2011, Chandra:PRE2011, Chandra:PRL2013}  This interesting phenomenon known as ``flow reversal"  remains unsolved, and it is related to the magnetic field reversals in stars and planets.  The properties of flow reversals depend quite critically on the box geometry, boundary conditions, and parameters, such as the Rayleigh number ($Ra$), which is the ratio of buoyancy and viscous forces, and the Prandtl number ($Pr$), which is the ratio of kinematic viscosity and thermal diffusivity. In this paper, we will investigate the properties of flow reversals in two-dimensional convective flow with free-slip boundary condition. 

Sugiyama \textit{et al.}~\cite{Sugiyama:PRL2010} performed a convection experiment with water in two quasi two-dimensional (2D) boxes of size $24.8 \mathrm{cm} \times 7.5 \mathrm{cm} \times 25.4 \mathrm{cm}$ and  $12.6 \mathrm{cm} \times 3.8 \mathrm{cm} \times 12.6 \mathrm{cm}$ under no-slip boundary condition,  and observed flow reversals for a range of Rayleigh and Prandtl numbers.  They observed a similar behavior in their numerical simulations. Both simulations and experiments show a  large-scale diagonal roll and two corner rolls.  The direction of the flow in the diagonal roll gets flipped after a reversal. Sugiyama \textit{et al.}~\cite{Sugiyama:PRL2010} attribute the flow reversal to the growth of the corner rolls due to plume detachments from the boundary layers.  Chandra and Verma~\cite{Chandra:PRE2011, Chandra:PRL2013} provided a quantitative description by showing that the flow reversals in the above 2D box  are related to  the nonlinear interactions among the large-scale structures.  During a reversal, the amplitude of the primary  mode vanishes, while the amplitudes of the secondary modes, especially the corner rolls, increase sharply.   

Breuer and Hansen~\cite{Breuer:EPL2009} simulated Rayleigh B\'{e}nard convection (RBC) in a two-dimensional box of aspect ratio two under free-slip boundary condition, and observed flow reversals for  infinite Prandtl number.  Here, the  flow profile is dominated by a single roll, which is represented by  the mode $(1,1)$.  The mode $(1,1)$ and the corresponding single roll flips during a flow reversal.   Petschel \textit{et al.}~\cite{Petschel:PRE2011} observed that several large-scale modes, namely modes $(1,1)$, $(2,1)$, and $(3,1)$, play an active role during a flow reversal.  

It is interesting to contrast the RBC flow structures for the no-slip and free-slip boundary conditions.  For the no-slip RBC simulation in a box of aspect ratio two, Chandra and Verma~\cite{Chandra:PRE2011} showed that the most dominant structures are  two horizontal rolls similar to the $(2,1)$ mode, and  two corner rolls.  Chandra and Verma~\cite{Chandra:PRE2011} showed that the corner rolls play an important role in the reversal dynamics.  In contrast, the free-slip simulations of Breuer and Hansen~\cite{Breuer:EPL2009} exhibit a large single roll, corresponding to the $(1,1)$ mode. The  $(2,2)$  mode is not the most dominant secondary mode for the free-slip RBC.   In the present paper, we perform a detailed analysis of the reversal dynamics for the free-slip RBC in boxes of the aspect ratio one and two, and contrast the reversal dynamics observed in the free-slip and no-slip boundary conditions.    We also compare the behavior of the Nusselt number for the two boundary conditions.

Free-slip motion is observed when a fluid moves over another fluid.   The Prandtl number of the mantle of the Earth is very large ($\sim 10^{25}$).  Hence, the flow of mantle over outer core in the Earth could be approximated by infinite Prandtl number RBC with free-slip boundary condition.    In addition, numerical simulations of RBC with the free-slip boundary condition is simpler compared to the no-slip boundary condition.  The  basis functions used for the free-slip boundary condition (composed of sin and cos functions)  are easier to analyse; these basis functions capture the  large-scale flow structures quite well, and provide valuable insights into the symmetries of the flow.  These simplifications are very useful for understanding the dynamics of flow reversals.

Large-scale structures play an important role in flow reversals in other geometries as well, for example, in a cylinder.   Brown \textit{et al.},~\cite{Brown:PRL2005} Brown and Ahlers,~\cite{Brown:JFM2006} and Xi and Xia~\cite{Xi:PRE2007, Xi:POF2008} studied flow reversals in a cylindrical geometry, and showed the importance of the large-scale circulation (LSC) in flow reversals. They  showed that the first mode vanishes abruptly during  a ``cessation-led reversals".    However, the ``rotation-led reversals" observed in a cylinder involves phase shifts of the dipolar mode.    Mishra \textit{et al.}~\cite{Mishra:JFM2011} performed numerical simulations in a cylinder of aspect ratio one and showed that for the cessation-led reversals, the dipolar mode decreases in amplitude, while the amplitude of the quadrupolar mode increases.  The nonlinear interactions during the cessation-led reversals in a cylinder have strong similarities with those in the two-dimensional box discussed earlier.  The interactions of the dipolar and quadrupolar modes in dynamo too show similar behavior,~\cite{Wicht:Geochem2004, Gallet:GAFD2012} thus making the study of nonlinear interactions among modes very important for the reversal studies.

Chandra and Verma~\cite{Chandra:PRE2011,Chandra:PRL2013} analysed the symmetries of flow reversals, and classified the modes that flip sign during a reversal.  In this paper, we extend their arguments and apply them to earlier simulations, as well as to the free-slip RBC simulation performed in the present paper.

The outline of the paper is as follows: in Sec.~\ref{sec:GE}, we discuss the  governing equations of the flow. In Sec.~\ref{sec:Sym}, we describe the symmetries of flow reversals.  Section~\ref{sec:DNS} contains the numerical method, while Sec.~\ref{sec:Rev} contains our results.  We conclude in the Sec.~\ref{sec:Con}.

\section{Governing equations}\label{sec:GE} 
In Rayleigh-B\'{e}nard convection, a Boussinesq fluid is placed between two horizontal plates separated by a distance $d$ and having a temperature difference $\Delta$. The RBC equations are
\begin{eqnarray}
\frac{\partial {\bf u}}{\partial t} + ({\bf u} \cdot \nabla){\bf u} & = & - \nabla \sigma + \alpha g \theta \hat{z} + \nu \nabla^2 {\bf u}, \label{eq:u} \\
\frac{\partial \theta}{\partial t} + ({\bf u} \cdot \nabla)\theta & = & -\frac{d\bar{T}}{dz} u_z +  \kappa \nabla^2 \theta, \label{eq:th} \\
\nabla \cdot {\bf u} & = & 0, \label{eq:cont_dimension}
\end{eqnarray}
where $\mathbf{u}$ is the velocity field, $\theta$ is the temperature fluctuations from the  conduction state,   $\sigma$ is the pressure field, $\nu,\kappa,\alpha$ are the kinematic viscosity, thermal diffusivity, and thermal expansion coefficient of the fluid, and $-g \hat{z}$ is the acceleration due to gravity.  Note that $d\bar{T}/dz = -\Delta/d$. We nondimensionalize the above equations using $\sqrt{\alpha g \Delta d/Pr}$ as the velocity scale, $d$ as the length scale, and $\Delta$ as the temperature scale. Hence the time scale used is the eddy turnover time, $d/\sqrt{\alpha g \Delta d/Pr}$. Throughout the paper,  time is referred to in the units of the eddy turnover time. The nondimensionalized RBC equations are
\begin{eqnarray}
\frac{1}{Pr}\left[ \frac{\partial {\bf u}}{\partial t} + ({\bf u} \cdot \nabla){\bf u} \right] & = & - \nabla \sigma + \theta \hat{z} + \frac{1}{\sqrt{Ra}} \nabla^2 {\bf u}, \label{eq:u_non} \\
\frac{\partial \theta}{\partial t} + ({\bf u} \cdot \nabla)\theta & = & u_z + \frac{1}{\sqrt{Ra}} \nabla^2 \theta, \label{eq:th_non} \\
\nabla \cdot {\bf u} & = & 0, \label{eq:cont}
\end{eqnarray}
The two nondimensionalized parameters are $Ra  = \alpha g \Delta d^3/\nu\kappa$ and $Pr = \nu/\kappa$. 
 
Under the limit of infinite Prandtl number, Eq.~(\ref{eq:u_non}) gets simplified to
\begin{equation}
- \nabla \sigma+  \theta \hat{z}+ \frac{1}{\sqrt{Ra}} \nabla^{2}\mathbf{u}  = 0,
\label{eq:NSE_infty}
\end{equation}
hence the momentum equation is linear in the $Pr=\infty$ limit.   For the analysis of  the large-scale flow structures and flow reversals, it is convenient to work in the Fourier space, in which the above equations for $Pr=\infty$ are 
\begin{eqnarray}
0 & = & - i \mathbf{k}\hat{\sigma}(\mathbf{k}) + \hat{\theta}(\mathbf{k}) \hat{z}  - \frac{1}{\sqrt{Ra}}  k^{2}\hat{\mathbf{u}}(\mathbf{k}) \label{eq:u_k}, \\
\frac{\partial \hat{\theta}(\mathbf{k})}{\partial{t}} & = & -ik_{j}\sum_{\mathbf{k}=\mathbf{p}+\mathbf{q}}\hat{u}_{j}(\mathbf{q}) \hat{\theta}(\mathbf{p}) + \hat{u}_z(\mathbf{k}) -\frac{1}{\sqrt{Ra}}  k^2\hat{\theta}(\mathbf{k}) \label{eq:theta_k}, \\
k_i\hat{u}_i(\mathbf{k})& = & 0,\label{eq:INC_k}
\end{eqnarray}
where $\hat{u}_i(\mathbf{k})$, $\hat{\theta}(\mathbf{k})$, and $\hat{\sigma}(\mathbf{k})$ are the Fourier transforms of the velocity, temperature, and pressure fields, respectively.  The equations in the Fourier space also reveals that in the $Pr=\infty$ limit, the nonlinearity in the system is present only in the temperature equation as $ -ik_{j} \hat{u}_{j}(\mathbf{q}) \hat{\theta}(\mathbf{p})$.  Using Eqs.~(\ref{eq:u_k},\ref{eq:INC_k}), Pandey \textit{et al.}~\cite{Pandey:PRE2014} derived the following relationships between the velocity and temperature modes:  
\begin{align}
\hat{u}_z(\mathbf{k})& = \sqrt{Ra}  \frac{k_{x}^2}{k^4}\hat{\theta}(\mathbf{k}),\label{eq:uz_k}\\
\hat{u}_x(\mathbf{k})& =  - \sqrt{Ra} \frac{k_zk_x}{k^4}\hat{\theta}(\mathbf{k}).\label{eq:ux_k}
\end{align}
Thus   $\hat{u}_{x,z}(\mathbf{k})$ are  proportional to  $\hat{\theta}(\mathbf{k})$, and hence, the velocity modes are slaved to the temperature modes. For finite but large $Pr$, the Fourier-transformed version of Eq.~(\ref{eq:u_non}) is
\begin{equation}
\frac{1}{Pr} \left[  \frac{\partial \hat{u}_i(\mathbf{k})}{\partial{t}}  + i k_{j} \sum_{\mathbf{k} \mathbf{p}+\mathbf{q}}\hat{u}_{j}(\mathbf{q}) \hat{u}_i(\mathbf{p})  \right] =  - i \mathbf{k}\hat{\sigma}(\mathbf{k}) + \hat{\theta}(\mathbf{k}) \hat{z}  - \frac{1}{\sqrt{Ra}}  k^{2}\hat{\mathbf{u}}(\mathbf{k})  \label{eq:u_k_finitePr}.
\end{equation}

We employ free-slip or stress-free boundary condition on all the four sides of the 2D box.  For the temperature field, we apply the conducting boundary condition ($\theta = 0$) at the top and bottom walls, and the insulating boundary condition ($\partial_x \theta = 0$) at the vertical walls.  For our simulations, we employ the following basis functions  that satisfies the boundary conditions:
\begin{eqnarray}
u_{x} & = & \sum_{k_x,k_z}  4 \hat{u}_{x}(k_x,k_z) \sin(k_{x}x)   \cos(k_{z}z)    \label{eq:ux_2d}, \\
u_{z} & = & \sum_{k_x,k_z} 4 \hat{u}_{z}(k_x,k_z) \cos(k_{x}x)  \sin(k_{z}z)  \label{eq:uy_2d}, \\
\theta & = & \sum_{k_x,k_z} 4 \hat{\theta}(k_x,k_z) \cos(k_{x}x) \sin(k_{z}z)  \label{eq:theta_2d}.
\end{eqnarray}
We refer to the above as {\em free-slip basis function}, for which we follow the conventions and definitions of FFTW.~\cite{Frigo:IEEE2005}   

In the next section, we discuss the symmetries of the convective flows; these symmetries provide valuable insights  into the flow reversals.

\section{Symmetries of the RBC equations and participating modes}\label{sec:Sym}
As described in the previous section, the time evolution of the velocity and temperature fields of a RBC  are given by Eqs.~(\ref{eq:u}$-$\ref{eq:cont_dimension}) in real space, and Eqs.~(\ref{eq:u_k}$-$\ref{eq:INC_k}, \ref{eq:u_k_finitePr}) in Fourier space. The  Eqs.~(\ref{eq:u}$-$\ref{eq:cont_dimension}) are invariant under $g \rightarrow -g$,  $d\bar{T}/dz \rightarrow -d\bar{T}/dz$, ${\mathbf u} \rightarrow {\mathbf u}$, and $\theta\rightarrow -\theta$. Physically, it corresponds to inverting the temperature gradient (putting the hot plate above, and the cold plate below) as well as the gravity.  Another symmetry in 2D is \{$x\rightarrow -x$; $u_x \rightarrow -u_x$\} which correspond to the mirror reflections perpendicular to the $x$.  In 3D, the corresponding symmetry along $y$ is \{$y \rightarrow -y$; $u_y \rightarrow -u_y$\}.  The reflection symmetry in 2D is also borne out by the basis function for $u_x$ given in Eq.~(\ref{eq:ux_2d}).

The solution of the RBC equations too show some interesting symmetry properties, which are of direct relevance to the reversal dynamics. Our simulation results show that some of the modes of Eqs.~(\ref{eq:ux_2d}$-$\ref{eq:theta_2d}) reverse sign, while  some do not.  In this section, we study the symmetry properties of these modes.  The modes of Eqs.~(\ref{eq:ux_2d}$-$\ref{eq:theta_2d}) belong to one of the four categories: even $E=$ (even, even), odd $O=$ (odd, odd), and mixed  $M_1= $ (even, odd), $M_2 = $ (odd, even).  $M_1$ and $M_2$  complement  each other, i.e., $\bar{M}_1 = M_2$ and $\bar{M}_2 = M_1$ under the operation ($\mathrm{even} \leftrightarrow \mathrm{odd}$).   To illustrate, $u_x(1,1)$ is an odd mode, $u_x(2,2)$ is an even mode, and $u_x(2,1)$ is a mixed mode of $M_1$ category.  

The nonlinear term of Eq.~(\ref{eq:theta_k})  is a sum of quadratic products of the modes.  If we focus on a unit  nonlinear interaction, then
\begin{equation} 
\partial_t \hat{\theta}(\mathbf k) \sim \hat{u}(\mathbf p)  \hat{\theta}(\mathbf q).
\label{eq:unit_triad}
\end{equation}
Here ${\bf k} = {\bf p} + {\bf q}$, which implies that $(n_x,n_z) =  (l_x +m_x, l_z+m_z)$, where $k_i = 2\pi n_i$, $p_i = 2\pi l_i$, and $q_i = 2\pi m_i$ with $i=(x,z)$ and $l_i, m_i, n_i$ as integers.  In free-slip basis [Eqs.~(\ref{eq:ux_2d}$-$\ref{eq:theta_2d})], however, a product of the modes $(m_1,n_1)$ and $(m_2,n_2)$ generate $(m_1\pm m_2, n_1 \pm n_2)$ modes; the $\pm$ in the resulting mode is due to the fact that $\sin(m_1 x)$ and $\cos(m_1 x)$  consist of $\exp( i m_1 x)$ and $\exp(- i m_1 x)$ modes.  Since even+even = even, even+odd = odd, and odd+odd = even, we obtain the product rules described in Table~\ref{tab:2d_product_rule}.  For example, $O\times O = E$, $O\times M_1 = M_2$.

\begin{table}
\caption{Rules of nonlinear interactions among the modes in RBC. The elements form the Klein four-group $Z_2 \times Z_2$.} 
\label{tab:2d_product_rule}
\begin{tabular}{|a | c | c | c | c|}
\hline
\rowcolor{Gray}
$\times$ & $E$ & $M_1$ & $M_2$ & $O$ \\ \hline
$E$ & $E$ & $M_1$ & $M_2$ & $O$ \\ \hline
$M_1$ & $M_1$ & $E$ & $O$ & $M_2$ \\ \hline
$M_2$ & $M_2$ & $O$ & $E$ & $M_1$ \\ \hline
$O$ & $O$ & $M_2$ & $M_1$ &  $E$ \\ \hline
\end{tabular}
\end{table}

The aforementioned four elements form an abelian group called {\em Klein four-group}, which is a direct product of two cyclic groups of two elements each, i.e., $Z_2 \times Z_2$.  There is a simple binary representation of this group: $(0,1) \times (0,1) = (00, 01, 10, 11)$, and the correspondence is $(E=00, M_1=01, M_2=10, O=11)$.  The $\mathrm{even}$ mode index is represented by $0$, and the $\mathrm{odd}$ index is  represented by $1$.

In a steady-state RBC flow, the modes typically fluctuate around a mean value, which could be finite or zero.  After a flow reversal, some of the modes  flip, i.e., their mean value changes sign.  Using the aforementioned product rules, we can discover which modes change sign after a reversal.  Table~\ref{tab:2d_product_rule} indicates that the properties of the table remain invariant for the combinations: $\{E, -O, M_1=\epsilon, M_2=\epsilon \}$, $\{E,  -M_1, O=\epsilon, M_2=\epsilon \}$, $\{E, -M_2, O=\epsilon, M_1=\epsilon,\}$, $\{E, -O, -M_1, M_2 \}$, $\{E, -O, M_1, -M_2 \}$, $\{E, O, -M_1, -M_2 \}$, where $\epsilon$ denotes fluctuating modes with zero mean.  The group structure also indicates that the modes of the class $E$, which is identity element of the group, can never change sign.   Thus, the rules for the change of sign of the modes  can be classified into  six classes:
\begin{enumerate}
\itemsep0em
\item 
$ \{ O \} \rightarrow  \{ -O \}$; $ \{ E \} \rightarrow  \{ E\}$; $ \{ M_1, M_2 \} =\epsilon$
\item 
$ \{ M_1 \} \rightarrow  \{ -M_1 \}$; $ \{ E \} \rightarrow  \{ E\}$; $ \{ O, M_2 \} =\epsilon$
\item 
$ \{ M_2 \} \rightarrow  \{ -M_2 \}$; $ \{ E \} \rightarrow  \{ E\}$; $ \{ O, M_1 \} =\epsilon$
\item
$ \{ O \} \rightarrow  \{ -O \}$; $ \{ M_1 \} \rightarrow  \{ -M_1 \}$; $ \{ M_2 \} \rightarrow  \{ M_2\}$; $ \{ E \} \rightarrow  \{ E\}$
\item
$ \{ O \} \rightarrow  \{ -O \}$; $ \{ M_2 \} \rightarrow  \{ -M_2 \}$; $ \{ M_1 \} \rightarrow  \{ M_1\}$; $ \{ E \} \rightarrow  \{ E\}$
\item
$ \{ M_1 \} \rightarrow  \{ -M_1 \}$; $ \{ M_2 \} \rightarrow  \{ -M_2 \}$; $ \{ O \} \rightarrow  \{ O\}$; $ \{ E \} \rightarrow  \{ E\}$
\end{enumerate}
 In the first three cases,  one class among $ \{ O, M_1, M_2 \}$ changes sign, while the other two classes are negligible.  For the latter three classes, two out of the three members  of the group $ \{ O \}, \{ M_1 \},\{ M_2 \}$ change sign after a reversal, and the third one remains unchanged.  The identity element $\{ E\}$ does not change sign.  Note that the above set of rules are generalizations of those described in Chandra and Verma.~\cite{Chandra:PRE2011}

We apply the above symmetry classes to several reversal works of the past.  In Table~\ref{tab:past_work_symmetry} we list the dominant modes and the dominant symmetry class of the numerical experiments of Chandra and Verma~\cite{Chandra:PRE2011} and van der Poel \textit{et al.}~\cite{Poel:PRE2011}, which are  for the no-slip boundary condition.  The simulation results of the present paper belongs to the same group as that of Breuer and Hansen.~\cite{Breuer:EPL2009}, which will be discussed in Sec.~\ref{subsec:AR2}.  Numerical simulations of van der Poel \textit{et al.}~\cite{Poel:PRE2011}  for aspect ratios $\Gamma=1/2,~1/4$ indicate dominance of $(1,2)$ and $(1,4)$ modes respectively, which belong to the class $\{ M_2 \}$.  If flow reversal takes place for the flow structure of van der Poel \textit{et al.}~\cite{Poel:PRE2011}, then we expect the modes to follow  rule $\hat{v}_{1,2} \rightarrow -\hat{v}_{1,2}$ for $\Gamma=1/2$.  Hence, the transformation rule for $\Gamma=1/2$ could belong to the rules (3), (5), or (6) listed above, depending on the strength of other members of group.   Hence, it will be interesting to analyse the modes of the flow in $\Gamma=1/2$ and $1/4$ boxes.   

\begin{table}
\begin{ruledtabular}
\caption{\label{jfonts}Classification of some of the 2D RBC systems.  The symmetry classification of van der Poel \textit{et al.}'s~\cite{Poel:PRE2011} RBC flows requires detailed study of the modes.} \label{tab:symmetry}
\begin{tabular}{ccccc}
Box  & Dominant & Generated  &Transformations & Symmetry \\
Geometry & modes & modes & during reversal & class\\
\hline
Chandra and Verma: $\Gamma=1$ & (1,1), (2,2) & $O,E$ & $\hat{v}_{1,1} \rightarrow -\hat{v}_{1,1}; \hat{v}_{2,2} \rightarrow \hat{v}_{2,2}$ & (1) \\
Chandra and Verma: $\Gamma=2$ & (2,1), (2,2)  & $M_1, E$ &  $\hat{v}_{2,1} \rightarrow -\hat{v}_{2,1}; \hat{v}_{2,2} \rightarrow \hat{v}_{2,2}$ & (2) \\
van der Poel \textit{et al.}: $\Gamma=1/2$ & (1,2), (2,2) & $M_{2}, E$ & $-$ & $-$ \\	
van der Poel \textit{et al.}: $\Gamma=1/4$ & (1,4),? & $M_{2}, E$ & $-$ & $-$ \\											
\end{tabular}
\label{tab:past_work_symmetry}
\end{ruledtabular}
\end{table}

In the next section, we will present our numerical method.

\section{Simulation details}\label{sec:DNS}
We perform numerical simulations of two-dimensional turbulent convection using a pseudo-spectral solver TARANG.~\cite{Verma:Pramana2013}    We employ the fourth-order Runge-Kutta (RK4) scheme for time advancement, Courant-Friedrichs-Lewy (CFL) condition for choosing the variable time step, and 2/3 rule for dealising. For our simulations, we choose two aspect ratios: $\Gamma=2$ and $\Gamma=1$ with $512\times 256$ and $512\times 512$ grid points respectively.   For the velocity field, we employ a free-slip boundary condition on all the walls, but for the temperature field,   we assume the top and bottom walls to be perfectly conducting, and the side walls to be perfectly insulating.  

We perform numerical simulations for $Ra$ ranging from $10^4$ to $10^8$ for both $\Gamma=1$ and $\Gamma=2$. We choose $Pr=\infty$ for which reversals are easier to obtain under the free-slip boundary condition. We also observe flow reversals for $Pr= 20$ and $40$, but we will not describe these results in detail.   We employ random initial condition for the simulation of $Ra=10^4$, and then we use the steady-state profile of the lower $Ra$ runs as an initial condition for the higher $Ra$ simulations.

We also perform a no-slip RBC simulation for $Pr=1$ in an aspect ratio two box to contrast the flow reversals in free-slip and no-slip boundary condition.  A brief detail of the no-slip  simulation is provided in Sec.~\ref{sec:no-slip}.   In the next section, we will discuss in detail the reversal dynamics for $\Gamma = 1$ and $2$ boxes.

\section{Dynamics of flow reversals}\label{sec:Rev}
Among all the simulations that we perform for $Pr=\infty$ and $Ra$ ranging from $10^4$ to $10^8$, we observe flow reversals for $Ra=10^7$ and $10^8$ in $\Gamma = 2$ box, and for $Ra = 10^8$ in $\Gamma=1$ box.  In the next subsection, we will study dynamics of these reversals.

\subsection{Flow reversals in a $\Gamma=2$ box for $Pr=\infty$} \label{subsec:AR2}
\label{sec:gamma_2}
We analyze the steady state data of our simulation for $Ra = 10^7$.  At first, we compute the most energetic  velocity modes.    In Table~\ref{tab:Gamma_2_modes}, we list the top 21 modes in a decreasing order of the modal kinetic energy $E_u(\mathbf k) = \langle |{\mathbf u}({\mathbf k})|^2 \rangle/2 $ during one of the flow reversals.  The value listed in the table is the average value of 400 eddy turnover time during a flow reversal.  Using Eqs.~(\ref{eq:uz_k},\ref{eq:ux_k}), we derive a relationship between the modal kinetic energy $E_u(\mathbf k)$ and the modal entropy $E_\theta(\mathbf k)$ as
\begin{equation}
E_u(m,n) = \frac{1}{2} \left(|\hat{u}_x(\mathbf k)|^2 + |\hat{u}_z(\mathbf k)|^2 \right) = \frac{1}{2}  Ra \frac{k_x^2}{k^6} |\hat{\theta}(\mathbf k)|^2 = Ra \frac{k_x^2}{k^6} E_\theta(\mathbf k) ,
\label{eq:Eu_Etheta}
\end{equation}
 where $k_x = m \pi/\Gamma$, $k_z = n\pi$, and $k^2 = k_x^2 + k_z^2$.  The modes $\hat{\theta}(0,2n)$ have an approximate amplitude of $-1/(2n\pi)$, as predicted by Mishra and Verma.~\cite{Mishra:PRE2010}

\begin{table}
\begin{ruledtabular}
\caption{For the free-slip RBC with $\Gamma=2$, $Pr=\infty$, and $Ra=10^7$, the most energetic 21 modes active during a flow reversal.  We average the modal kinetic energy $E_u({\mathbf k}) = |\hat{u}({\mathbf k})|^2/2$ for 400 eddy turnover time during a reversal.}
\label{tab:A2-T_Modes}
\begin{tabular}{cc|cc|cc}
$(m,n)$ & $E_u({\mathbf k}) = |\hat{u}({\mathbf k})|^2/2$ & $(m,n)$ & $E_u$ & $(m,n)$ & $E_u$\\
\hline
$(1,1)$ & $9.27 \times 10^{-2}$ & $(6,1)$ & $8.10 \times 10^{-4}$ & $(3,3)$ & $2.56 \times 10^{-4}$ \\
$(3,1)$ & $1.61 \times 10^{-2}$ & $(3,2)$ & $7.42 \times 10^{-4}$ & $(7,2)$ & $2.28 \times 10^{-4}$ \\
$(2,1)$ & $9.90 \times 10^{-3}$ & $(4,2)$ & $6.68 \times 10^{-4}$ & $(7,3)$ & $1.78 \times 10^{-4}$ \\
$(4,1)$ & $3.54 \times 10^{-3}$ & $(5,2)$ & $4.83 \times 10^{-4}$ & $(4,3)$ & $1.74 \times 10^{-4}$ \\
$(5,1)$ & $3.31 \times 10^{-3}$ & $(1,2)$ & $4.02 \times 10^{-4}$ & $(6,3)$ & $1.31 \times 10^{-4}$ \\
$(7,1)$ & $1.08 \times 10^{-3}$ & $(6,2)$ & $3.60 \times 10^{-4}$ & $(1,3)$ & $1.20 \times 10^{-4}$ \\ 
$(2,2)$ & $8.28 \times 10^{-4}$ & $(5,3)$ & $2.70 \times 10^{-4}$ & $(2,3)$ & $9.49 \times 10^{-5}$ \\ 
\end{tabular}
\label{tab:Gamma_2_modes}
\end{ruledtabular}
\end{table}

\begin{figure}
\begin{center}
\includegraphics[scale=0.4]{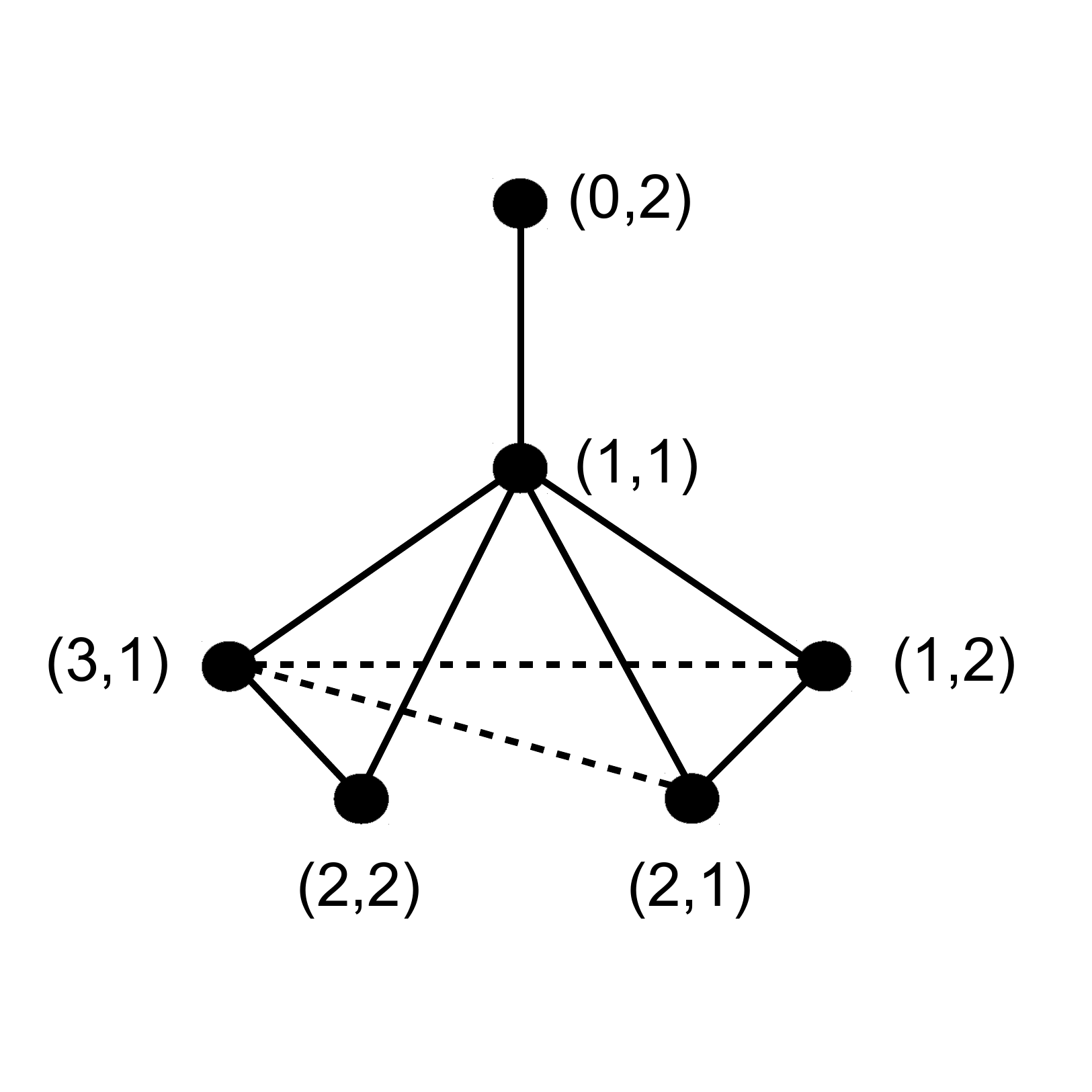}
\caption{Important modes and triadic interactions for free-slip RBC during a flow reversal.  The mode (1,1) has the maximum energy.  We illustrate some of the most dominant interacting triads, which are $\{ (1,1), (2,2), (3,1) \}$, $\{ (3,1), (2,1), (1,2) \}$,  and  $\{ (1,1), (1,2), (2,1) \}$.  The mode $\hat{\theta}(0,2)$, generated by the triad $\{ \hat{\mathbf u}(1,1), \hat{\theta}(1,1), \hat{\theta}(0,2) \}$, does not participate directly in the dynamics of flow reversals.}
\label{fig:triads}
\end{center}
\end{figure}

The nonlinear interactions of the temperature equation [Eq.~(\ref{eq:theta_k})] involves triad interactions among $\{\hat{u}(\mathbf q), \hat{\theta}(\mathbf p),  \hat{\theta}(\mathbf k) \}$ (two $\theta$ modes and one $u$ mode) with ${\mathbf p +\mathbf q = \mathbf k}$.~\cite{Mishra:PRE2010}    A large number of wavenumber triads  participate in nonlinear interactions, but we focus our attention on triads $\{ (1,1), (2,2), (3,1) \}$, $\{ (3,1), (2,1), (1,2) \}$,  and  $\{ (1,1), (1,2), (2,1) \}$, which are some of the most dominant triad interactions during a flow reversal (see Fig.~\ref{fig:triads}).   The physical interpretation of the participating modes are as follows --- $(1,1)$:  a single convective roll;  $(2,2)$: four rolls in a $2\times 2$ grid; $(1,2)$ and $(2,1)$: two rolls stacked along $y$ and $x$ directions respectively; $(3,1)$: three rolls stacked along $x$.  The triad $\{ (\hat{\theta}(0,2), \hat{\theta}(1,1), \mathbf{\hat{u}}(-1,1) \}$ is also important in RBC,~\cite{Mishra:PRE2010} but it is not very critical for a flow reversal.   Refer to Mishra and Verma~\cite{Mishra:PRE2010} for an interpretation and importance of $\hat{\theta}(0,2)$ mode in RBC. 

\begin{figure}
\begin{center}
\includegraphics[scale=0.35]{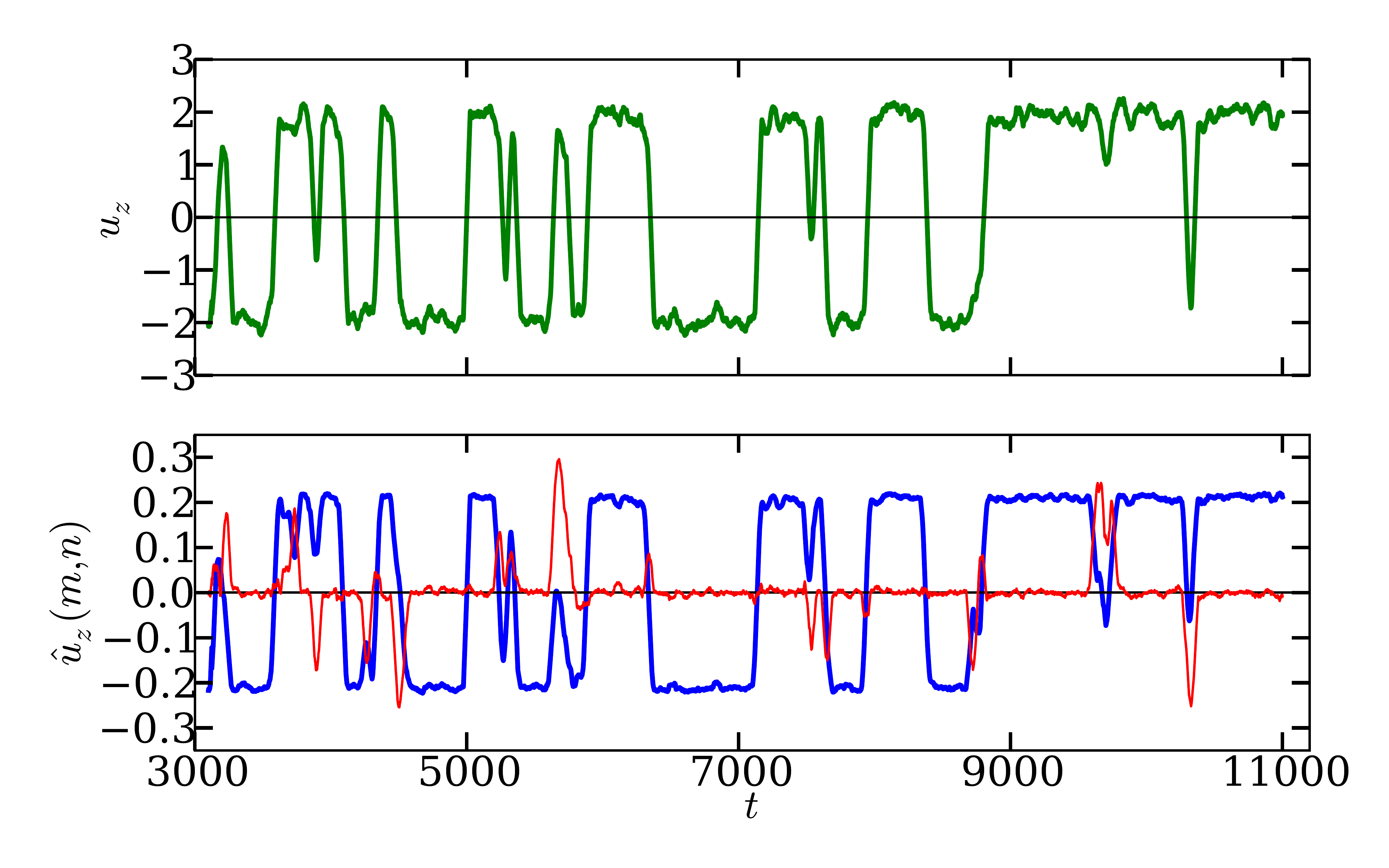}
\caption{For RBC simulation with  free-slip boundary condition and parameters $\Gamma = 2, Ra = 10^7$, and  $Pr=\infty$: (top panel) the time series of $u_{z}$ at a location $(x=0.01,z=0.50)$; (bottom panel) the time series of the amplitude of the modes $\hat{u}_{z}(1, 1)$ and $\hat{u}_{z}(2, 1)$, represented by the blue  and red curves respectively. Time in this plot and subsequent plots are in units of eddy turnover time.}
\label{fig:v_real}
\end{center}
\end{figure}

\begin{figure}
\includegraphics[scale=0.3]{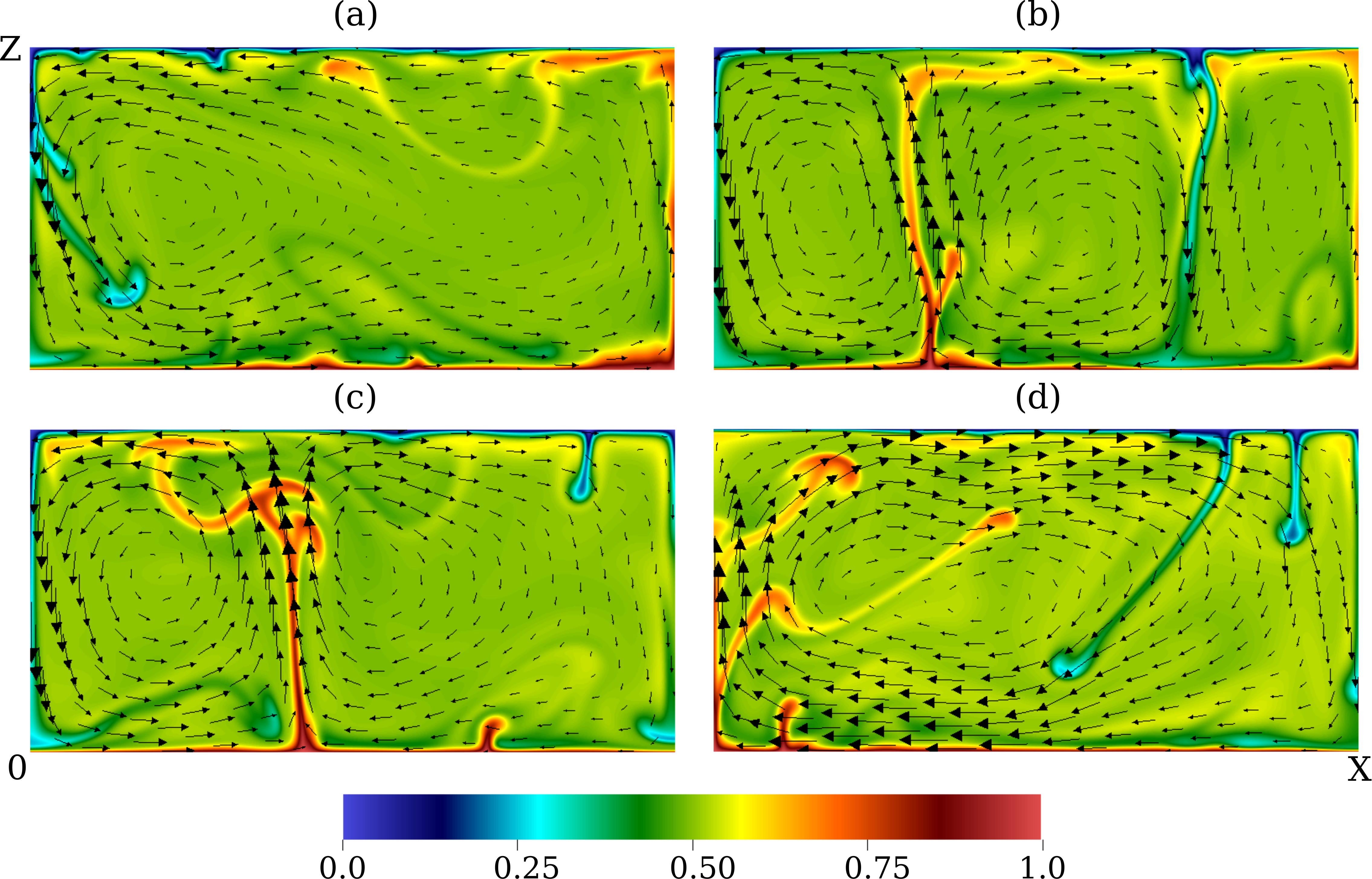}
\caption{For RBC simulation with  free-slip boundary condition and parameters $\Gamma = 2, Ra = 10^7$, and  $Pr=\infty$: the snapshots of the temperature and velocity fields at $t_a=3120.0$, $t_b=3150.0$,  $t_c=3162.0$, and $t_d=3190.0$ exhibiting a flow reversal.  The blue and red colors depict  the coldest and hottest regions, respectively.  The black arrows represent the velocity field. }
\label{fig:profile_AR2}
\end{figure}

In Fig.~\ref{fig:v_real}(a), we plot the time series of the vertical velocity measured at the real space probe located at $(x=0.01,z=0.50)$, which is near the centre of the left wall.  In Fig.~\ref{fig:v_real}(b), we  plot the time series of the amplitude of the modes $\hat{u}_z(1,1)$  and $\hat{u}_z(2,1)$. The time series exhibits large fluctuations;  to smoothen the plots, we perform a running average of the real space and modes time series over 101 data points, that is, $\langle f(i) \rangle = (\sum_{i-50\le j \le i+50}  f(j) )/101$.  The vertical velocity at the probe and the mode $\hat{u}_z(1,1)$ exhibit reversals, indicating that the large-scale circulation in the box reverses during  flow reversals. The flow profiles (velocity and temperature fields) before and after one of the reversals are exhibited  in Fig.~\ref{fig:profile_AR2}(a) and Fig.~\ref{fig:profile_AR2}(d) respectively; they demonstrate the dominance of a single roll structure, represented by the $(1,1)$ mode.

When we zoom in one of the reversals, we observe an interesting dynamics between the flow structures.  Figure~\ref{fig:profile_AR2} exhibits four snapshots (a,b,c,d), in which the blue color represents the coldest regions, while the red color represents the hottest regions.  The velocity fields are shown using arrows.  The snapshot (a) exhibits a dominant single role, consistent with the prominent $(1,1)$ mode of the time series. The snapshot  (b) contains a three-roll structure, which corresponds to  the $(3,1)$ mode, along with the $(1,1)$ mode.  In snapshot (c), the mode $(2,1)$, corresponding to the two roll structure, is  most dominant.  Finally, in the snapshot (d), the intermediate mode $(2,1)$  weakens, and the mode $(1,1)$ again becomes strong, but with a reversed sign.   Note that the transition from snapshot (b) to (c) of Fig.~\ref{fig:profile_AR2} involves deletion of the right-most roll of (b), while the transition  from (c) to (d) involves the deletion of the left roll.   The change of sign of the $(1,1)$ mode leads to a reversal of the vertical velocity.  

\begin{figure}
\includegraphics[scale=0.35]{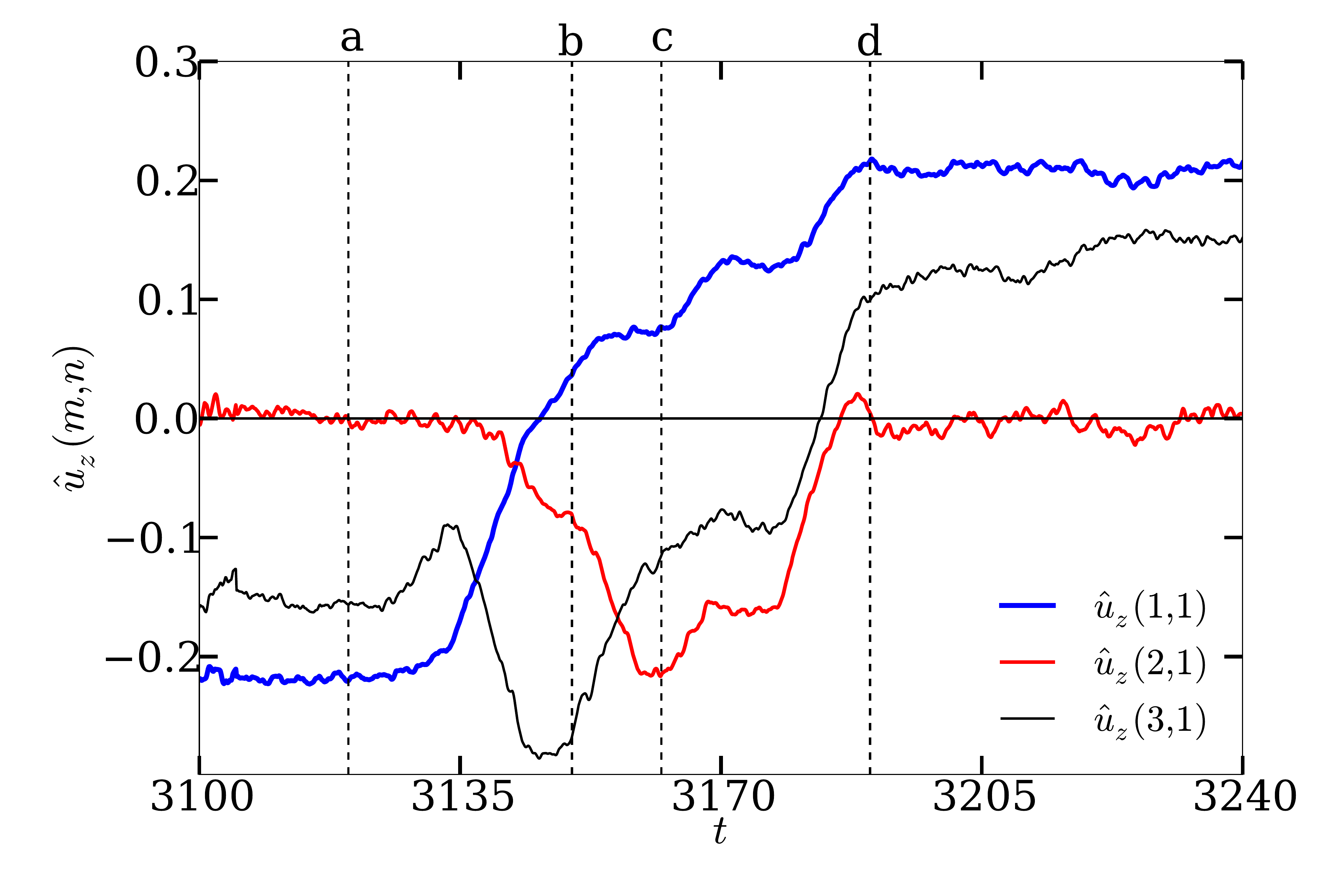}
\caption{For RBC simulation with  free-slip boundary condition and parameters $\Gamma = 2, Ra = 10^7$, and  $Pr=\infty$: the time series of the amplitude of dominant modes near the flow reversal whose snapshots are shown in Fig.~\ref{fig:profile_AR2}. The times at a, b, c, and d correspond to four snapshots shown in Fig.~\ref{fig:profile_AR2}. }
\label{fig:time_series_snapshots}
\end{figure}

\begin{figure}
\includegraphics[scale=0.3]{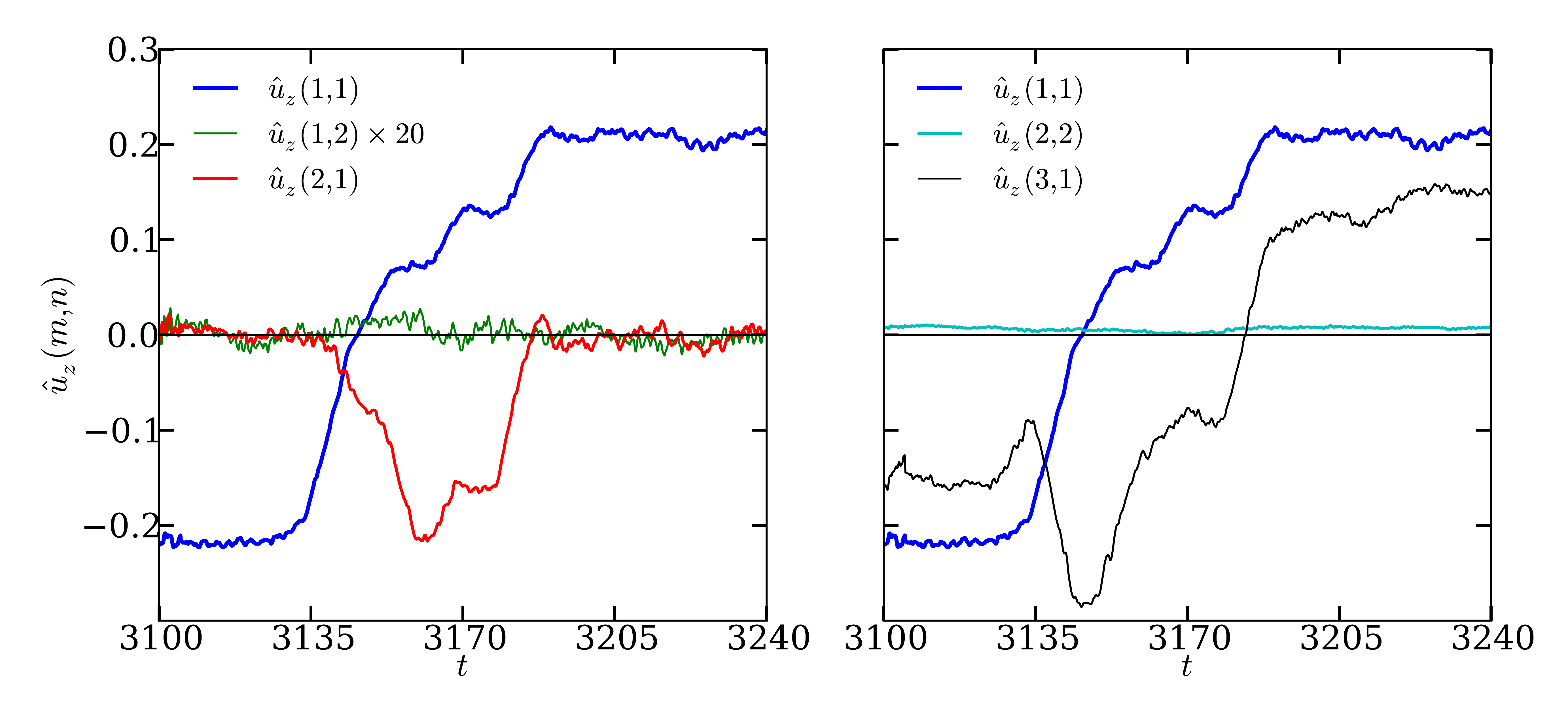}
\caption{For RBC simulation with  free-slip boundary condition and parameters $\Gamma = 2, Ra = 10^7$, and  $Pr=\infty$: time series of the amplitudes  of some of the  dominant modes.}
\label{fig:triad_time_series}
\end{figure}

The time series of the amplitude of the modes during the aforementioned flow reversal is shown in Fig.~\ref{fig:time_series_snapshots}, in which the vertical lines $a$ to $d$ represent the times of the snapshots (a$-$d) of Fig.~\ref{fig:profile_AR2}, respectively.  Figure~\ref{fig:time_series_snapshots} shows that the mode $(3,1)$ dominates in snapshot (b), but vanishes in (c). The mode $(2,1)$ remains dominant from (c) to (d); beyond (d), the mode $(2,1)$ vanishes, and the mode $(1,1)$ becomes dominant.  The sign of $(1,1)$ changes from (a) to (d).  Time series of the amplitude of some of the other dominant modes during the flow reversal, $(1,1), (2,1), (3,1), (1,2)$,  $(2,2)$,  are shown in Fig.~\ref{fig:triad_time_series} (also see Table~\ref{tab:A2-T_Modes}).  

A closer observation of the flow structures reveals that during a reversal, the intermediate mode (3,1), corresponding to three rolls, is enhanced first via a triad interaction among $\{ (1,1), (3,1), (2,2) \}$.  After this, the mode $(2,1)$, corresponding to two rolls, increases in amplitude via another triad interaction $\{ (3,1), (2,1), (1,2) \}$. The transition from the two-roll structure to the single roll structure of Fig.~\ref{fig:time_series_snapshots} involves a triad interaction $\{ (2,1), (1,1), (1,2) \}$.  Interestingly, the intermediate roll (2,2) is much weaker than the other modes (see Table~\ref{tab:A2-T_Modes}), in sharp contrast to the dominant role played by the (2,2) mode in the no-slip RBC (see Chandra and Verma~\cite{Chandra:PRE2011, Chandra:PRL2013}).

The signs of the dominant modes show interesting pattern.  We observe that after a reversal, all the odd modes flip, but the even ones retain their sign.  The mixed modes ($M_1$ and $M_2$ of Sec.~\ref{sec:Sym}) are quite insignificant (except during a reversal) for this geometry.  Therefore, we conclude that in a flow reversal in RBC with the free-slip boundary condition, the $O$ modes change sign, $E$ modes do not change sign, and $M_1$ and $M_2$ modes are insignificant.  Thus, the reversals for $\Gamma=2$ box belong to class (1) listed in Sec.~\ref{sec:Sym}.

\begin{figure}
\includegraphics[scale=0.25]{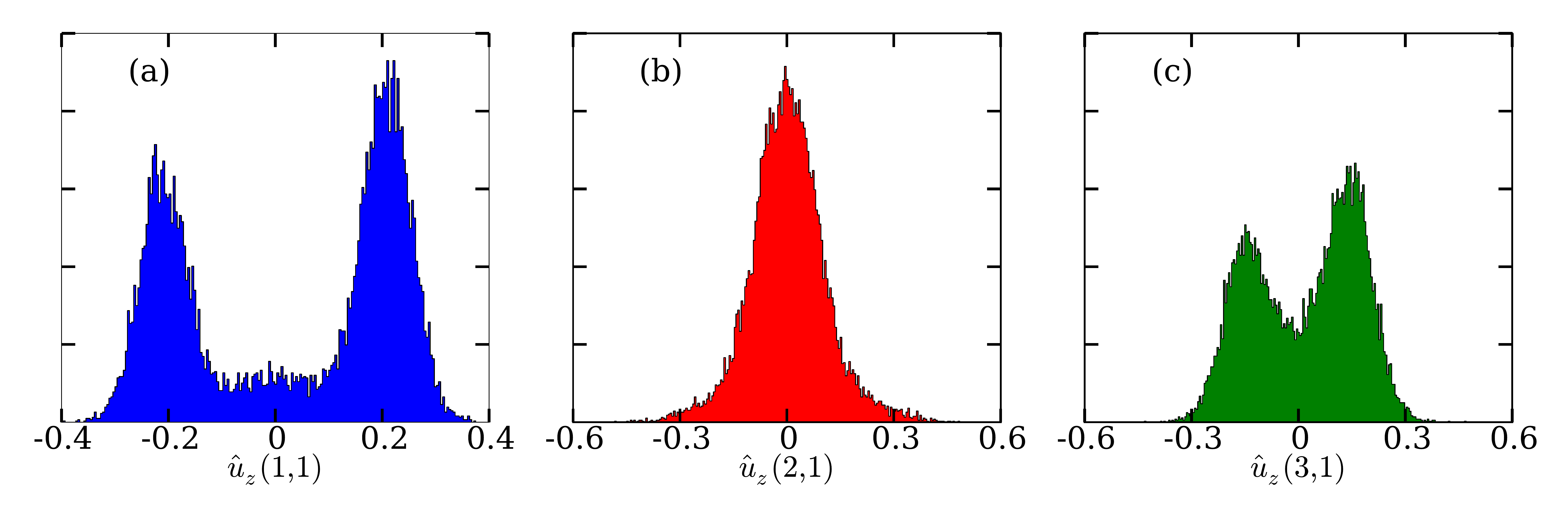}
\includegraphics[scale=0.25]{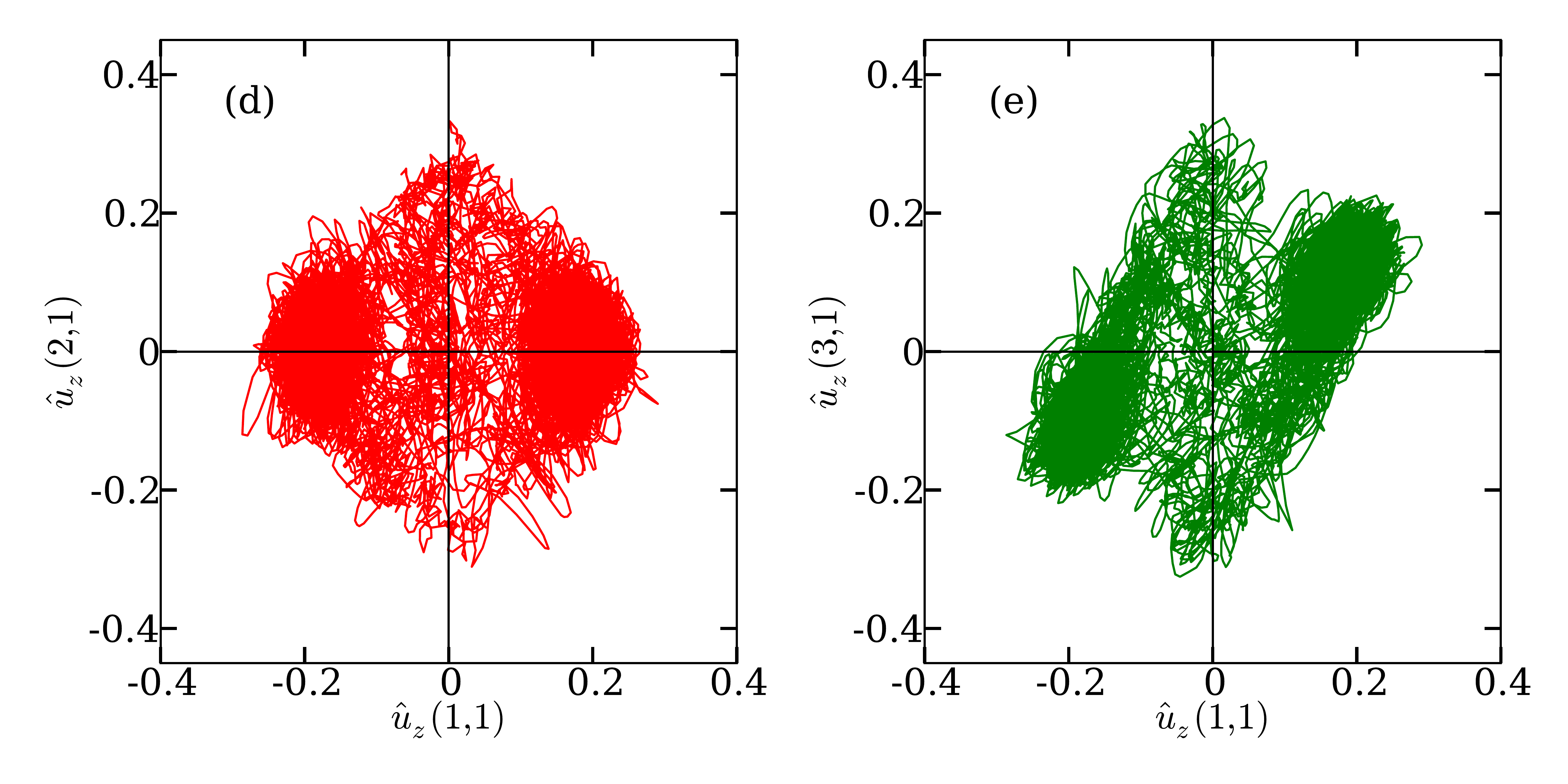}
\caption{For RBC simulation with  free-slip boundary condition and parameters $\Gamma = 2, Ra = 10^7$, and  $Pr=\infty$: the probability distribution function of some of the dominant modes --- (a) $\hat{u}_z(1,1)$, (b) $\hat{u}_z(2,1)$, and (c) $\hat{u}_z(3,1)$; phase space plots of (d) $\hat{u}_z(2,1)$ vs. $\hat{u}_z(1,1)$, and (e) $\hat{u}_z(3,1)$ vs. $\hat{u}_z(1,1)$}
\label{fig:A2-PDF-Phase}
\end{figure}

The above results are also borne out in the probability density functions (PDF) of $\hat{u}_z(1,1)$, $\hat{u}_z(2,1)$, and $\hat{u}_z(3,1)$, shown in Fig.~\ref{fig:A2-PDF-Phase}.  The double hump of  $\hat{u}_z(1,1)$ and $\hat{u}_z(3,1)$ illustrates switching of their signs during a reversal.  The  PDF of  $\hat{u}_z(2,1)$ mode indicates that this mode fluctuates around zero. Note that the PDF of $\hat{u}_z(3,1)$ mode should be symmetric if we perform our simulations for much longer time; the asymmetry in the figure is purely due to limited time span of the simulation.  These results are consistent with the phase space plots shown in Fig.~\ref{fig:A2-PDF-Phase}(d,e).   The dense regions in the phase space plots represent the non-reversing regions, while the fluctuations in the phase space illustrate the dynamics during a reversal; the modes exhibit significant fluctuations during a flow reversal.  We observe dense region for nonzero  $\hat{u}_z(1,1)$, $\hat{u}_z(3,1)$, but for $\hat{u}_z(2,1) \approx 0$, which is consistent with the PDF results that  $\hat{u}_z(1,1)$ and $\hat{u}_z(3,1)$ have nonzero mean value, but $\hat{u}_z(2,1)$ fluctuates around zero.  Our PDF and phase space results are consistent with those of Petschel \textit{et al.}~\cite{Petschel:PRE2011}

In the next subsection we compare our free-slip results with the reversal dynamics in RBC with the no-slip boundary condition.

\subsection{Comparison with flow reversals in no-slip RBC}
\label{sec:no-slip}
We performed RBC simulations for the same geometry, but for $Pr=1$ and $Ra=10^7$, and with no-slip boundary condition for the velocity field on all the walls.  For the temperature field we employ conducting boundary condition at the top and bottom walls, but insulating boundary condition at the side walls.  The simulations were performed using NEK5000~\cite{Fischer:JCP1997} that uses spectral element method. We used a $48\times 28$ spectral elements along with a seventh order polynomial, with higher resolutions near the boundaries.  See Chandra and Verma~\cite{Chandra:PRE2011} for more details.  

\begin{figure}
\includegraphics[scale=0.45]{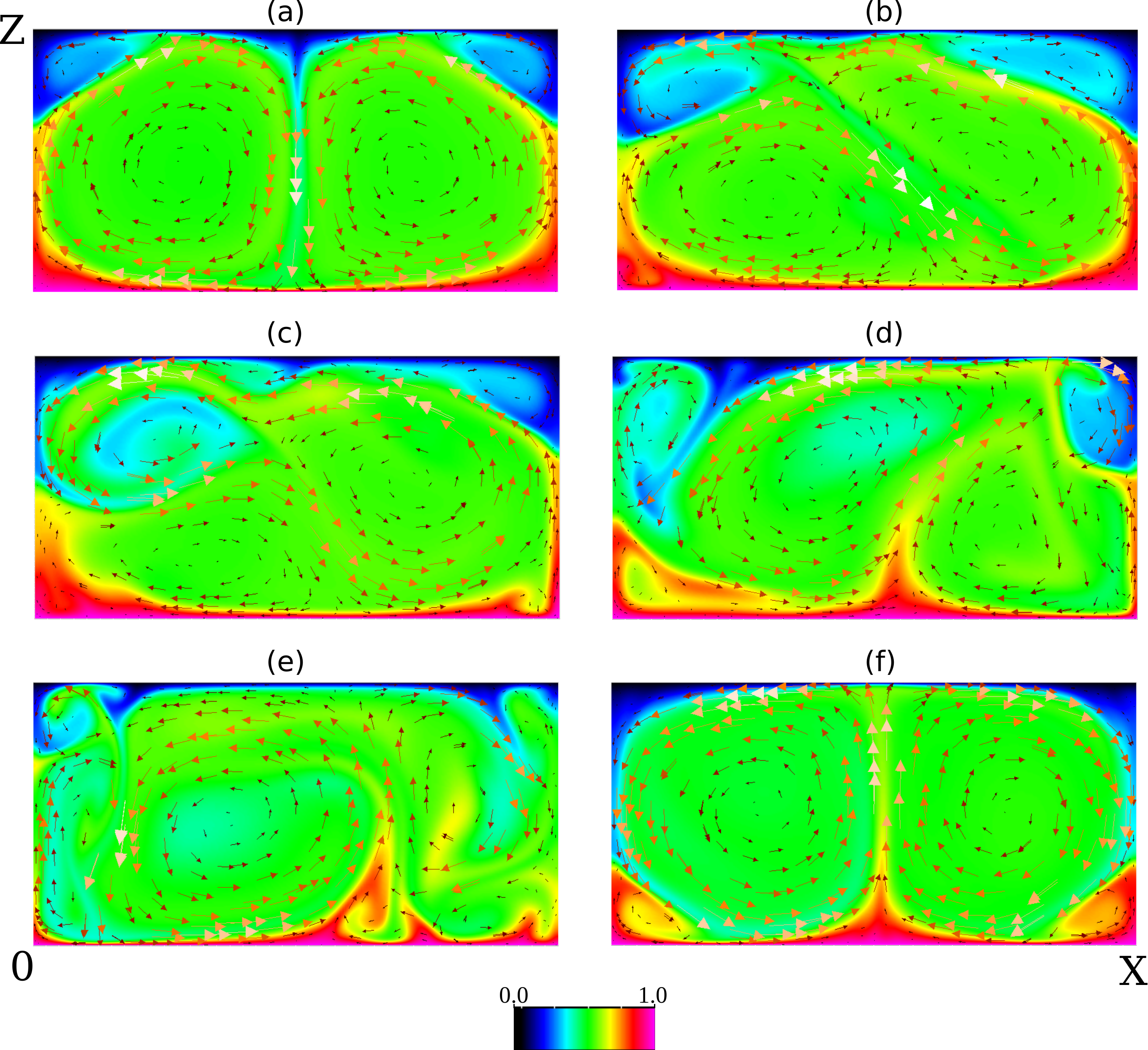}
\caption{For RBC simulation with  no-slip boundary condition and parameters $\Gamma = 2, Ra = 10^7$, and  $Pr=1$: six snapshots exhibiting a flow reversal and vortex reconnection (see Chandra and Verma~\cite{Chandra:PRL2013} for $\Gamma=1$).}
\label{fig:no_slip}
\end{figure}

The flow reversals in a no-slip box occur via a vortex reconnection, as reported by Chandra and Verma~\cite{Chandra:PRL2013} for a $\Gamma=1$ box.  Here, we briefly describe the flow reversal dynamics for the $\Gamma=2$ box.  Six snapshots of the velocity and temperate fields during a flow reversal are shown in Fig.~\ref{fig:no_slip} with the same color convention as Fig.~\ref{fig:profile_AR2}.     In Fig.~\ref{fig:no_slip}(a), we observe two large rolls, and two corner rolls near the top plate.  The top-left and the bottom-right rolls turn counterclockwise, while  the other two rolls turn clockwise.  In the early phase of a flow reversal, the left corner roll  grows in size, as shown in  Fig.~\ref{fig:no_slip}(b).  At a later time, the flow configuration appears as snapshot (c), in which the top-left corner roll and the bottom-right roll come closer and reconnect, and form a large vortex.\cite{Chandra:PRL2013}   Subsequently, the large  vortex
moves to the left, and the bottom-left roll gets squeezed and moves to the right, as shown in  snapshot (d). The two dominant rolls reorganize as shown in snapshots (d), (e), and (f).  The final configuration, shown in snapshot (f), contains two large rolls, and two corner rolls  near the bottom plate.  The large rolls of snapshot (f)  have velocity fields opposite  to that of snapshot (a). The vortex reconnection in the above description is similar to that observed by Chandra and Verma~\cite{Chandra:PRL2013} for the no-slip RBC in a square box.  

A comparison between the dynamics of flow reversals between the  free-slip and no-slip boundary conditions reveal that for the free-slip boundary condition, the corner rolls and vortex reconnection are absent during a flow reversal, in contrast to active role played by them in flow reversals for the no-slip boundary condition.  Also, the dominant structures for the boundary conditions are very different; a single roll for the free-slip, but two rolls for the no-slip boundary condition.

\begin{figure}
\includegraphics[scale=0.57]{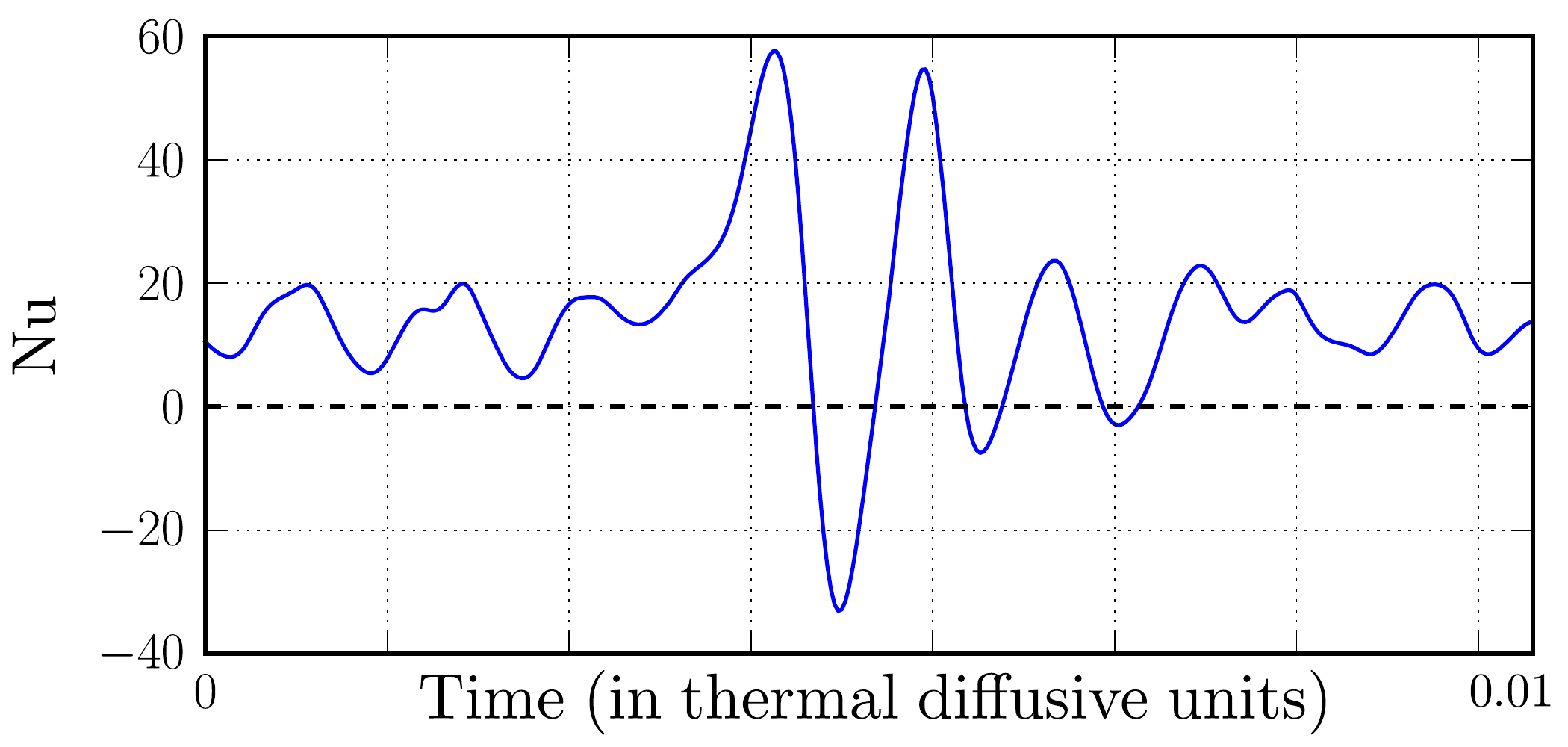}
\caption{The time series of $Nu$ near a flow reversal for $\Gamma = 1$ with no-slip boundary condition on all walls. The governing parameters are $Pr = 1$ and $Ra = 2 \times 10^7$. Reprinted with permission from M. Chandra and M. K. Verma, Phys. Rev. Lett. 110, 114503 (2013). Copyright 2013, American Physical Society.}
\label{fig:Nu_no_slip}
\end{figure}

\begin{figure}
\includegraphics[scale=0.35]{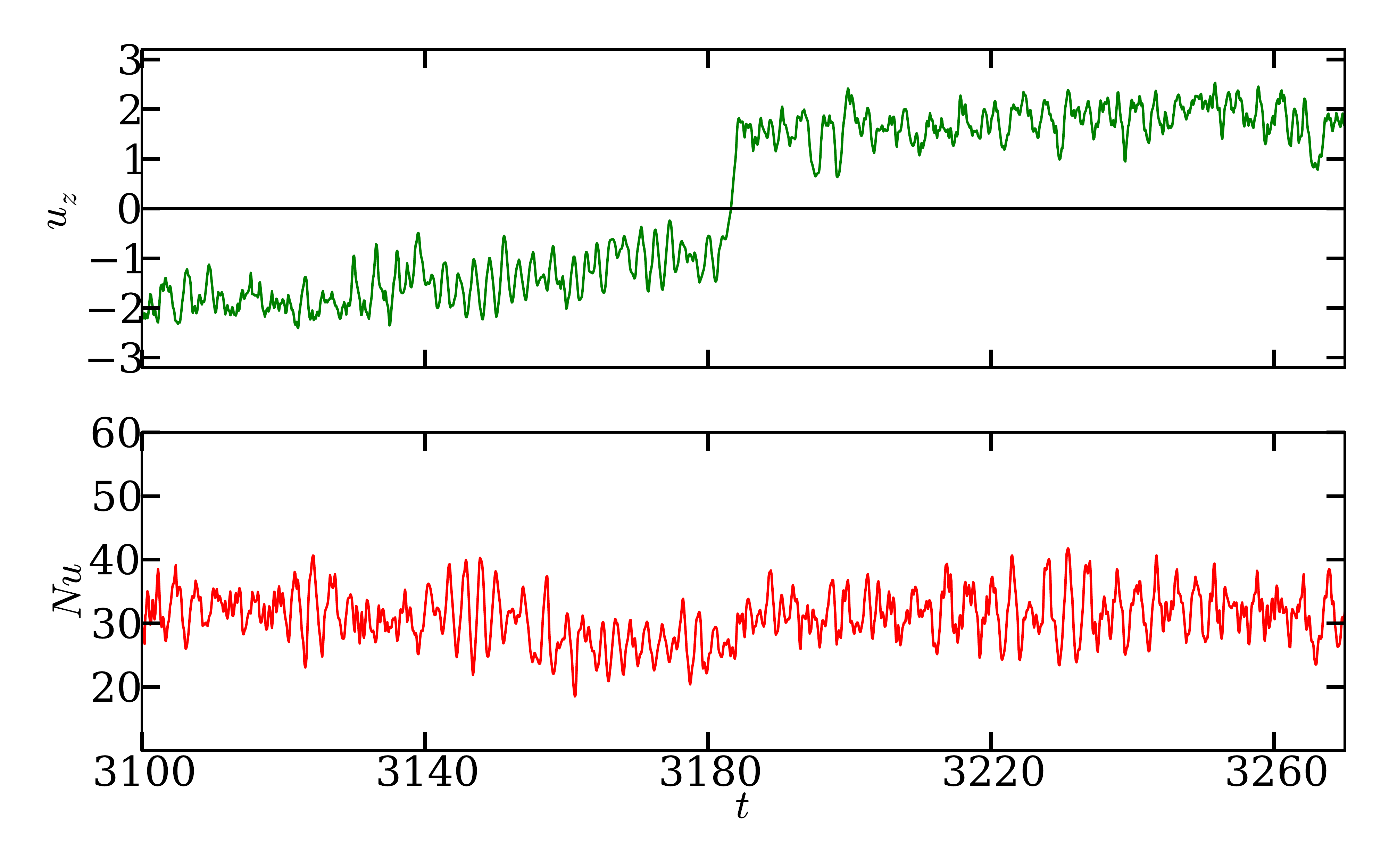}
\caption{For RBC simulation with  free-slip boundary condition and parameters $\Gamma = 2, Ra = 10^7$, and  $Pr=\infty$: the time series of $u_z$ (at a probe near the left wall) and $Nu$ near a flow reversal.}
\label{fig:Nu}
\end{figure}

The Nusselt number ($Nu$), which is a ratio of the total heat transfer and the conductive heat transfer, is an important quantity in RBC. Chandra and Verma~\cite{Chandra:PRE2011, Chandra:PRL2013} reported strong fluctuations in Nusselt number for no-slip RBC (see Fig.~\ref{fig:Nu_no_slip}).  However, the fluctuations in $Nu$ for the free-slip RBC is  comparatively much weaker, as shown in Fig.~\ref{fig:Nu}.  This difference is related to the weak $(2,2)$ mode in the free-slip RBC.

In the next subsection, we will present simulation results for a $\Gamma=1$ box with the free-slip boundary condition.

\subsection{Flow reversals in a $\Gamma=1$ box for $Pr = \infty$}
We also performed simulations of RBC flow in a 2D box of unit aspect ratio with free-slip boundary condition on all the walls.  The governing parameters for the simulation are $Pr = \infty$ and $Ra = 10^8$. The dominant velocity modes during a reversal (averaged over 400 eddy turnover time) are listed in Table~\ref{tab:Gamma_1_modes}. Like $\Gamma=2$, the three most dominant modes participating in the flow  reversal are $(1,1)$, $(3,1)$, and $(2,1)$.  Some of the most dominant nonlinear triads are $\{ (1,1), (2,2),  (3,1) \}$, $\{ (3,1), (2,1), (2,2) \}$, and $\{ (1,1), (1,2), (2,2) \}$, similar to  $\Gamma=2$ case. 

\begin{table}
\begin{ruledtabular}
\caption{For RBC simulation with free-slip boundary condition and parameters $\Gamma=1$, $Pr = \infty$, and $Ra = 10^8$, the most energetic 21 modes active during a flow reversal.  We average the modal kinetic energy $E_u({\mathbf k}) = |\hat{u}({\mathbf k})|^2/2$ for 400 eddy turnover time during a reversal.}
\begin{tabular}{cc|cc|cc}
$(m,n)$ & $E_u({\mathbf k}) = |\hat{u}({\mathbf k})|^2/2$  & $(m,n)$ & $E_u$ & $(m,n)$ & $E_u$ \\
\hline
$(1,1)$ & $2.12 \times 10^{-1}$ & $(1,2)$ & $5.26 \times 10^{-4}$ & $(6,1)$ & $1.10 \times 10^{-4}$ \\
$(3,1)$ & $2.03 \times 10^{-2}$ & $(5,3)$ & $5.26 \times 10^{-4}$ & $(5,5)$ & $1.08 \times 10^{-4}$ \\
$(2,1)$ & $6.07 \times 10^{-3}$ & $(4,1)$ & $4.53 \times 10^{-4}$ & $(4,2)$ & $8.91 \times 10^{-5}$ \\
$(5,1)$ & $3.09 \times 10^{-3}$ & $(2,2)$ & $4.52 \times 10^{-4}$ & $(2,3)$ & $8.53 \times 10^{-5}$ \\
$(3,3)$ & $1.18 \times 10^{-3}$ & $(3,2)$ & $2.78 \times 10^{-4}$ & $(4,3)$ & $7.19 \times 10^{-5}$ \\
$(7,1)$ & $8.82 \times 10^{-4}$ & $(7,3)$ & $2.01 \times 10^{-4}$ & $(7,5)$ & $5.87 \times 10^{-5}$ \\ 
$(1,3)$ & $8.50 \times 10^{-4}$ & $(3,5)$ & $1.39 \times 10^{-4}$ & $(5,2)$ & $5.29 \times 10^{-5}$ \\
\end{tabular}
\label{tab:Gamma_1_modes}
\end{ruledtabular}
\end{table}

\begin{figure}
\includegraphics[scale=0.9]{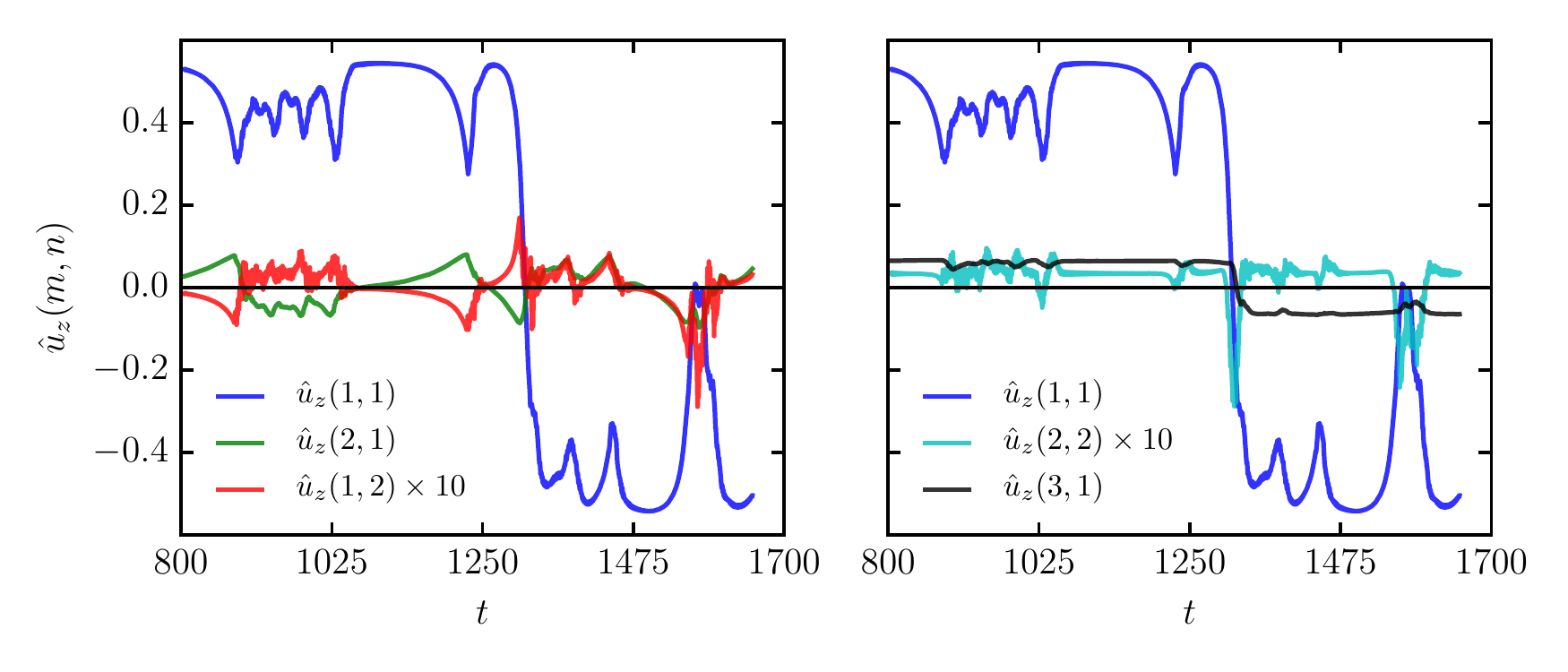}
\caption{For RBC simulation with  free-slip boundary condition and parameters $\Gamma = 1, Ra = 10^8$, and  $Pr=\infty$: the time series of the amplitudes  of some of the dominant modes during a flow reversal.}
\label{fig:mode_time_gamma_1}
\end{figure}

\begin{figure}
\includegraphics[scale=0.31]{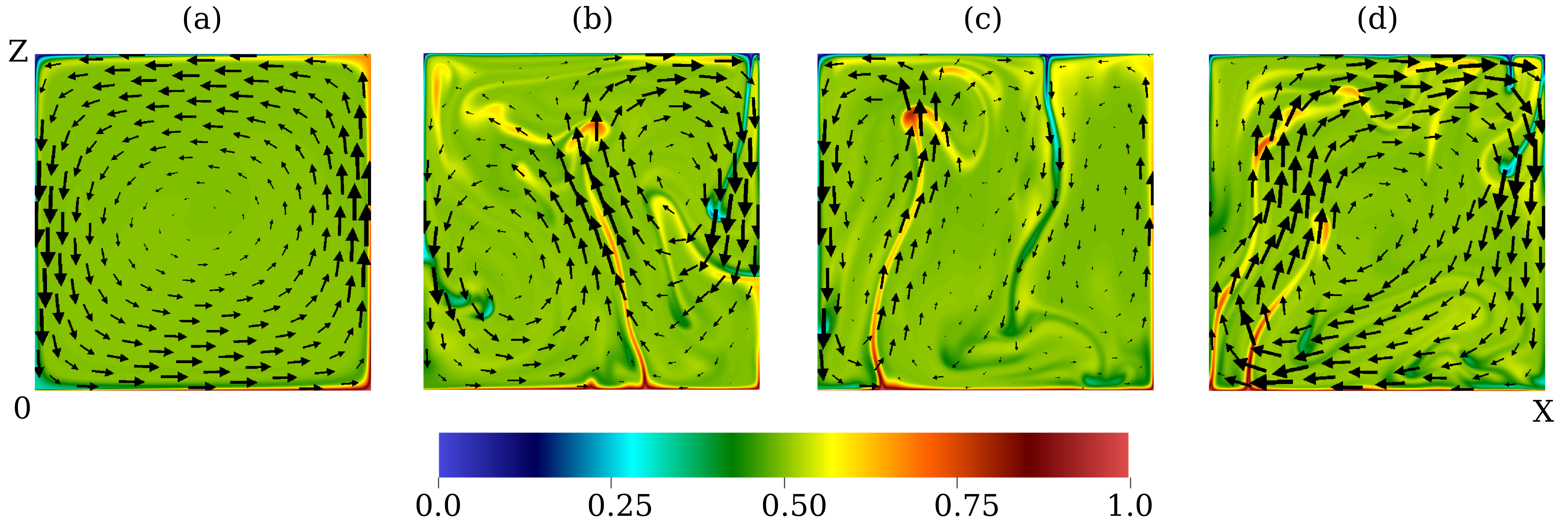}
\caption{For RBC simulation with  free-slip boundary condition and parameters $\Gamma = 1, Ra = 10^8$, and  $Pr=\infty$:  the snapshots of the temperature and velocity fields at $t_a=1100.0$, $t_b=1316.7$,  $t_c=1324.0$, and $t_d=1328.0$ exhibiting a flow reversal. Same color convention as Fig.~\ref{fig:profile_AR2}.}
\label{fig:profile_AR1}
\end{figure}

In Fig.~\ref{fig:mode_time_gamma_1}, we plot the time series of the amplitude  of the interacting modes.  We observe that the odd modes $(1,1)$ and $(3,1)$ reverse sign after the reversal, while the mode $(2,2)$ does not change sign except briefly near the reversal.  The modes $(2,1)$ and $(1,2)$ fluctuate about zero.  The flow profiles during the reversal, shown in Fig.~\ref{fig:profile_AR1},  have similarities with those for $\Gamma=2$.  Figures~\ref{fig:profile_AR1}(b) and~\ref{fig:profile_AR1}(c) contain two and three rolls, corresponding to the $(2,1)$ and $(3,1)$ modes, respectively.  After these intermediate rolls, the flow reorganizes itself as a dominant single roll, but with the sense of rotation opposite to the original one.

\subsection{Flow reversals in a $\Gamma=2$ box for $Pr=20,40$}
We also simulate free-slip RBC  in a box of aspect ratio $\Gamma=2$ for a set of Prandtl and Rayleigh numbers.  We observe flow reversals for $Ra = 10^8$ with $Pr=20$ and 40, whose dynamics is similar to that described in Sec.~\ref{sec:gamma_2}.  No reversals were observed for $Pr < 20$ for Rayleigh number up to $10^8$.  We are not certain why free-slip RBC exhibit flow reversals for large and infinite Prandtl numbers only, unlike no-slip RBC that shows reversals for $Pr=0.8$ to 10 and possibly beyond.~\cite{Sugiyama:PRL2010}     The properties of the flow reversals for $Pr=20$ and 40 have strong similarities with that for $Pr=\infty$, hence  we do not discuss them here.

\subsection{Summary of flow reversal dynamics for free-slip boundary condition}
\begin{figure}
\includegraphics[scale=0.31]{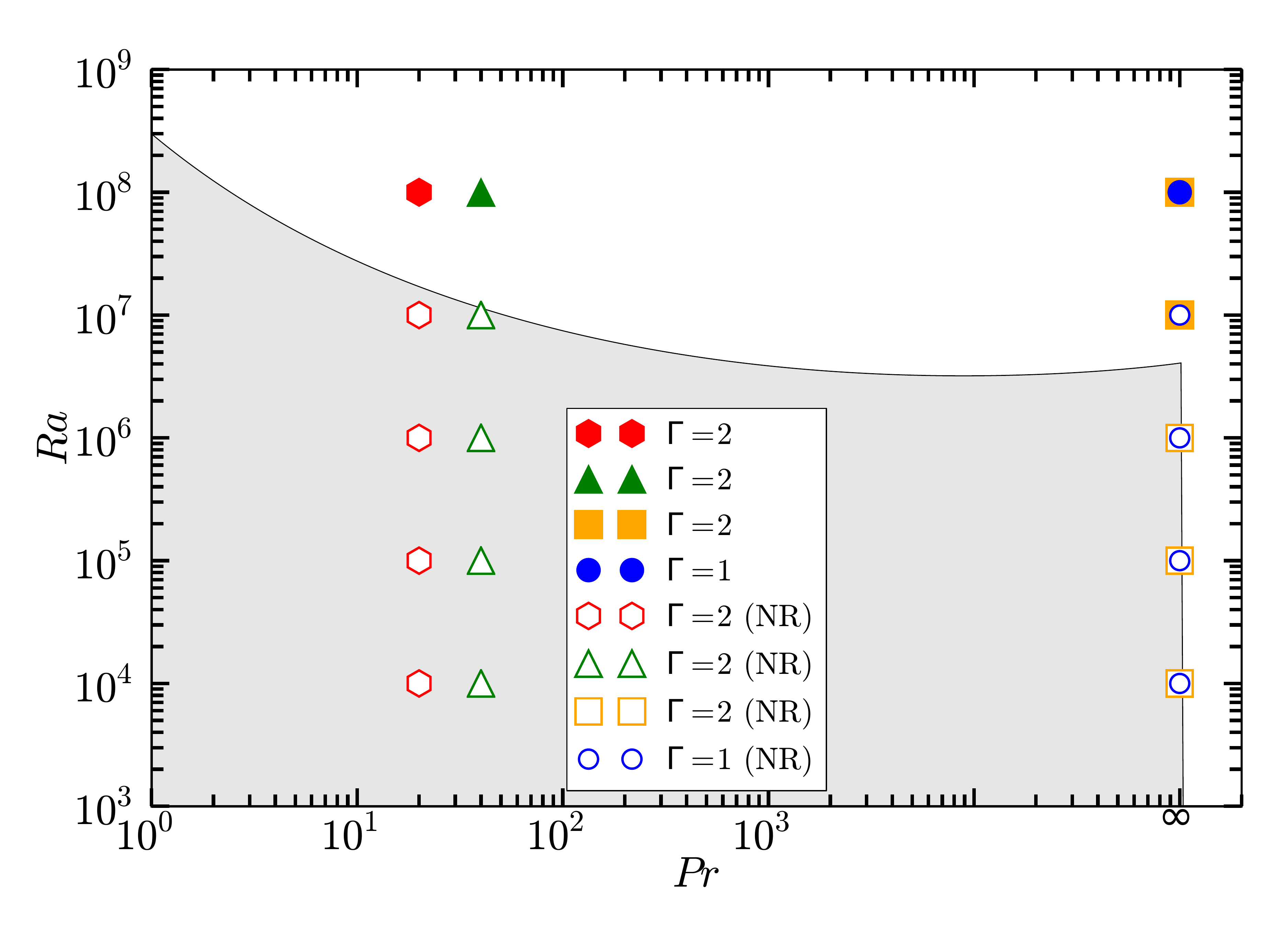}
\caption{For RBC simulations with  free-slip boundary condition: Parameter space plot of the reversal/non-reversal states. The filled symbols represent the parameters for which reversals occur, while unfilled ones represent parameters for which reversals do not occur.  We project that the flow reversal does not occur for the GREY region in the parameter space.}
\label{fig:phase}
\end{figure}

\begin{figure}
\includegraphics[scale=0.31]{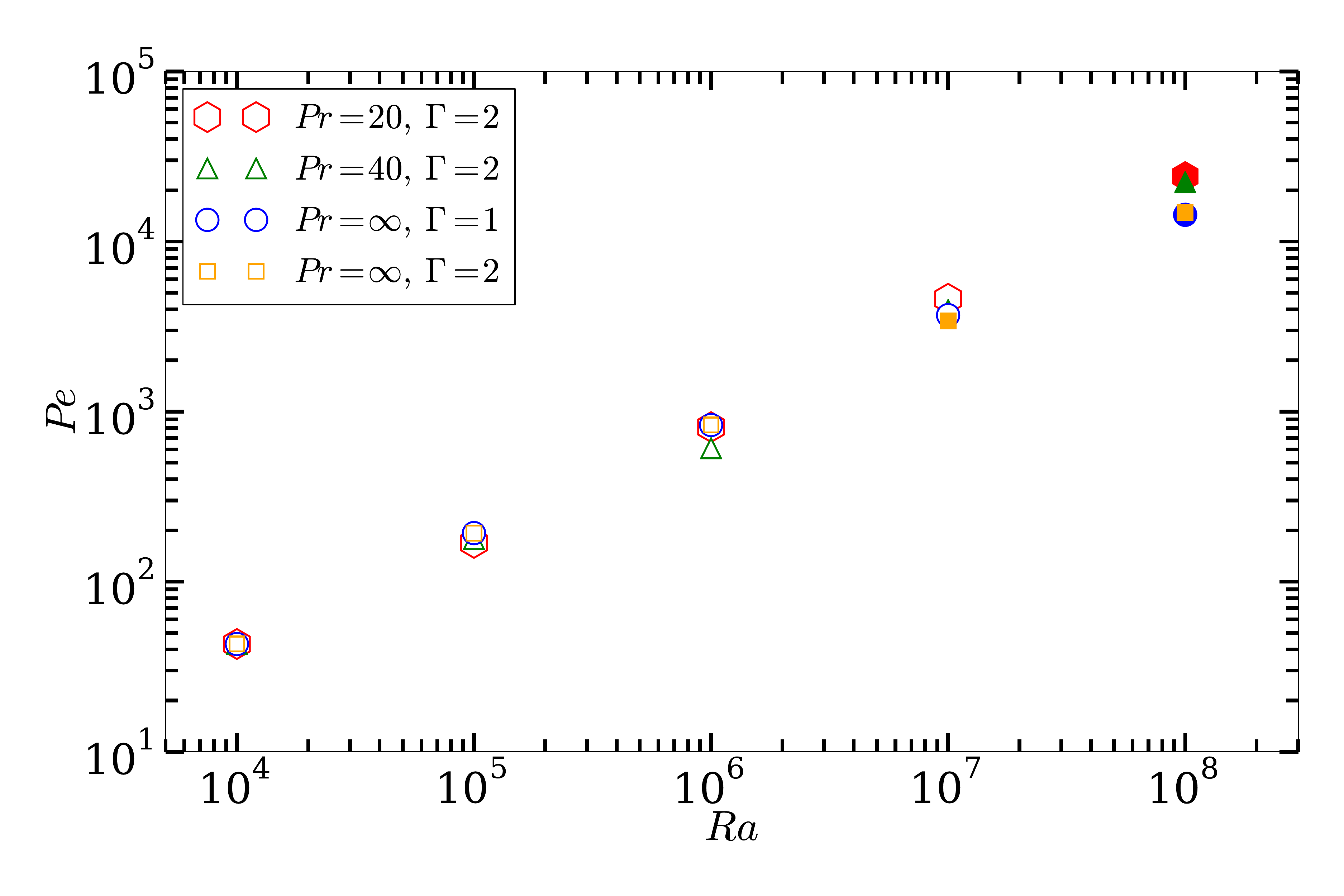}
\caption{For RBC simulations with  free-slip boundary condition: P\'{e}clet number as a function of Rayleigh number for various Prandtl numbers. The parameters with filled symbols show reversals, while those with unfilled ones do not show any reversal.}
\label{fig:Peclet}
\end{figure}

 In Fig.~\ref{fig:phase}, we  summarize  the parameter regimes that show flow reversals.  Figure~\ref{fig:Peclet} exhibits P\'{e}clet number for the runs performed by us.   We could perform simulations only for a limited range of parameters due to heavy computational cost.  The illustrated phase diagram provides useful insight into the reversal dynamics in geometries with free-slip boundary condition. 
 
The flow reversals with the free-slip boundary condition are typically observed at large Prandtl and Rayleigh numbers.   The Reynolds number of such flows is quite small, hence the momentum equation (Navier-Stokes equation) is linear.  However, the P\'{e}clet number, shown in Fig.~\ref{fig:Peclet}, is significantly large for such flows.\cite{Pandey:PRE2014}  Hence, in the temperature equation, the nonlinear term  is much larger than the diffusion term.  The flow reversals occur due to this nonlinearity.  In contrast, for no-slip boundary condition at moderate Prandtl numbers, the nonlinear term of the momentum equation, ${\bf u}\cdot \nabla {\bf u}$,  plays a major role during the flow reversals. For example, Sugiyama {\em et al.}~\cite{Sugiyama:PRL2010} and Chandra and Verma~\cite{Chandra:PRE2011,Chandra:PRL2013} showed that the flow reversals in two dimension for the no-slip boundary condition  typically  stop at very large Rayleigh numbers due to the strengthening of the large scale structures; this phenomena has been attributed to the inverse energy cascade of kinetic energy.    Note that the ${\bf u}\cdot \nabla {\bf u}$ term  is absent or weak in the flows with free-slip boundary condition at large Prandtl numbers.  Hence, the dynamics of flow reversals with no-slip and free-slip boundary conditions are quite different, with  ${\bf u}\cdot \nabla {\bf u}$ and  ${\bf u}\cdot \nabla \theta$ playing active roles for the respective boundary conditions.  This is the reason for the difference between the phase diagrams of ours and Sugiyama {\em et al.}~~\cite{Sugiyama:PRL2010}.

\section{Conclusions and Discussions}\label{sec:Con}
In this paper, we simulated 2D RBC with the free-slip boundary condition for aspect ratios one and two.   In Fig.~\ref{fig:phase} we sketch the parameter regime that shows flow reversals, and in Fig.~\ref{fig:Peclet} we plot the corresponding P\'{e}clet numbers.  We observe that the flow reversals are easier with the increase of Prandtl number. We did not observe flow  reversal for $Pr < 20$ with Rayleigh number up to $10^8$. For the free-slip boundary condition at large Prandtl numbers, the flow reversals  occur due to the nonlinearity ${\bf u}\cdot \nabla \theta$ of the temperature equation.  In contrast, for the no-slip boundary condition, the nonlinear term  ${\bf u}\cdot \nabla {\bf u}$ of the Navier Stokes plays a major role during a reversal.  At present we do not understand clearly why the flow reversals stop at low and moderate Prandtl numbers for the free-slip boundary condition.

Our numerical results show that the  modes  $(1,1), (2,1)$, and  $(3,1)$ play a dominant role during a flow reversal  for both the geometries.  The mode $(1,1)$ or a single roll is the most dominant large-scale flow structure.  During the reversal, the primary structure $(1,1)$ weakens, and the secondary modes $(3,1)$, $(2,1)$  become prominent.   Interestingly, $(1,1)$ and $(3,1)$ change sign,  the mode $(2,2)$ retains its sign, while $(2,1)$, and $(1,2)$ fluctuates around zero.   By performing a detailed analysis of these modes we deduce that $\{ E \} \rightarrow \{ E \}, \{ O \} \rightarrow \{ -O \}$, and $\{ M_1, M_2  \} \rightarrow 0$, which is the symmetry class (1) discussed in Sec.~\ref{sec:Sym}.  

The reversal dynamics of 2D RBC with free-slip boundary condition has certain similarities and dissimilarities with that of no-slip boundary condition. For both the boundary conditions, the flow reversals is intimately connected to the nonlinear interactions among the large-scale modes.   However, the corner rolls (part of a 4-roll structure, similar to (2,2) mode) plays a crucial role in no-slip RBC, whereas the $(3,1)$ and $(2,1)$  are the most important modes in the flow reversals for free-slip RBC.  The $(2,2)$ mode in free-slip is much weaker than the corresponding mode in the no-slip boundary condition.  Also, the Nusselt number fluctuations for the free-slip boundary condition is much weaker than that for the no-slip boundary condition.  In addition, the flow configurations under the no-slip and free-slip boundary conditions are different.  For example, for the no-slip RBC with $\Gamma=2$ and large $Ra$, a pair of rolls is the most dominant  flow structure,~\cite{Chandra:PRE2011}   but under the free-slip boundary condition, the flow is dominantly a large single roll structure.

We also present symmetry arguments to derive class of modes that could change sign during a flow reversal in arbitrary situation.  We show that the modes $\{ E \}, \{ O \}, \{ M_1 \}, \{ M_2 \}$ form a Klein four-group, which is a product of two cyclic groups $Z_2 \times Z_2$.  The above identification of the modes with one of the standard groups helps us in the classification of the reversing modes.  The above symmetry arguments can be easily generalized to higher dimensions.

Thus, our results confirm the importance of large-scale structures in flow reversals.  Similar arguments are applicable to other geometries like cylinder, cuboids, and spheres.  Still some intriguing questions remain unanswered:  why do we observe flow reversals only for large- and infinite Prandtl numbers for the free-slip boundary condition?  Why the aspect ratio one and two have very similar dynamics?  We are in process of constructing several low-dimensional models for this system, that we may clarify some of the above questions.

\section*{Acknowledgments}
We  thank Arpit Sahu for performing some set of initial runs, and Mani Chandra for sharing the data and plots of simulations with no-slip boundary condition. The simulations for $Pr=20$ were performed earlier by K. S. Reddy.  We thank Stephan Fauve for the valuable suggestions and references.  We also thank A. Kumar, B. Dutta, and A. G. Chatterjee for their valuable suggestions on matplotlib and programming.   This work was supported by a research grant SERB/F/3279/2013-14 from Science and Engineering Research Board, India, and CEFIPRA/4904.  Our numerical simulations were performed on {\em chaos} and {\em newton} clusters of IIT Kanpur.

\end{document}